\documentclass[11pt,letter]{article}
\usepackage{auxlib}
\usepackage{natbib}

\pdfoutput=1 

\usepackage{authblk}

\title{Nash Social Welfare with Submodular Valuations: Approximation Algorithms and Integrality Gaps}

\author[1]{Xiaohui Bei }
\author[2]{Yuda Feng \thanks{The work of Yuda Feng and Shi Li was supported by the State Key Laboratory for Novel Software Technology and the New Cornerstone Science Laboratory.}}
\author[3]{Yang Hu }
\author[2]{Shi Li $^\ast$}
\author[4]{Ruilong Zhang }

\affil[1]{\small Division of Mathematical Sciences, Nanyang Technological University, Singapore}
\affil[2]{\small School of Computer Science, 
Nanjing University, Nanjing, China}
\affil[3]{\small Institute for Interdisciplinary Information Sciences, Tsinghua University, Peking, China}
\affil[4]{\small Department of Mathematics, Technical University of Munich, Munich, Germany}

\affil[ ]{\texttt{xhbei@ntu.edu.sg, yudafeng@smail.nju.edu.cn, y-hu22@mails.tsinghua.edu.cn, shili@nju.edu.cn, ruilong.zhang@tum.de}}

\date{}

\begin{document}

\maketitle

\begin{abstract}
We study the problem of allocating items to agents with submodular valuations with the goal of maximizing the weighted Nash social welfare (NSW).  
The best-known results for unweighted and weighted objectives are the $(4+\epsilon)$ approximation given by Garg, Husic, Li, Végh, and Vondrák~[STOC 2023] and the $(233+\epsilon)$ approximation given by Feng, Hu, Li, and Zhang~[STOC 2025], respectively.

In this work, we present a $(3.56+\epsilon)$-approximation algorithm for weighted NSW maximization with submodular valuations, simultaneously improving the previous approximation ratios of both the weighted and unweighted NSW problems. 
Our algorithm solves the configuration LP of Feng, Hu, Li, and Zhang~[STOC 2025] via a stronger separation oracle that loses an $e/(e-1)$ factor only on small items, and then rounds the solution via a new bipartite multigraph construction. 
Some key technical ingredients of our analysis include a greedy proxy function, additive within each configuration, that preserves the LP value while lower-bounding the rounded solution, together with refined concentration bounds and a series of mathematical programs analyzed partly by computer assistance.

On the hardness side, we prove that the configuration LP for weighted NSW with submodular valuations has an integrality gap of at least $(2^{\ln 2}-\epsilon) \approx 1.617 - \epsilon$, which is slightly larger than the current best-known $e/(e-1)-\epsilon \approx 1.582-\epsilon$ hardness of approximation~[SODA 2020].
For additive valuations, we show an integrality gap of $(e^{1/e}-\epsilon)$, which proves the tightness of the approximation ratio in~[ICALP 2024] for algorithms based on the configuration LP. 
For unweighted NSW with additive valuations, we show an integrality gap of $(2^{1/4}-\epsilon) \approx 1.189-\epsilon$, again larger than the current best-known $\sqrt{8/7} \approx 1.069$-hardness of approximation for the problem~[Math. Oper. Res. 2024].
\end{abstract}

\newpage
\tableofcontents

\clearpage
\newpage

\section{Introduction}
\label{sec:intro}

We investigate a resource allocation problem that allocates a set $M$ of indivisible items among a set $N$ of agents.
Each agent $i\in N$ is associated with a monotone non-negative submodular valuation function $v_i: 2^M \to \R_{\geq 0}$ and a weight $w_i\in(0,1)$, where $\sum_{i\in N}w_i=1$.
The goal is to find an allocation $\cS:=(S_i)_{i\in N}$ that maximizes the weighted Nash Social Welfare (NSW), which is the weighted geometric mean of agents' valuations:
\[
\mathrm{NSW}(\cS)=\prod_{i\in N}(v_i(S_i))^{w_i}.
\]
A special case of this problem is the unweighted NSW, where $w_i=1/n$ for each $i$.

There are two traditional objectives for this resource allocation problem, focusing on efficiency and fairness of the allocation: the first \emph{utilitarian} approach maximizes the sum of individual utilities. This problem, known as the submodular welfare problem, admits an $e/(e-1)$-approximation~\cite{stoc/Vondrak08}; the \emph{egalitarian} approach maximizes the minimum of all individual utilities.
This is also known as the submodular Santa Claus problem, and the current best-known result is the $O(n^{\epsilon})$-approximation in $n^{O(1/\epsilon)}$-time given by~\cite{soda/BamasMR25}.
Unlike these two objectives, the Nash social welfare objective strikes a balance between fairness and efficiency, making it a central problem in algorithmic game theory, economics, and resource allocation. 
It has applications in bargaining theory~\cite{CM10,LV07,Tho86}, climate agreements~\cite{YvIW17}, and equitable division of resources~\cite{DWL+18,HdLGY14}.

As an important special case of submodular functions, the Nash social welfare problem is well-studied under additive valuations.
For unweighted NSW, Barman, Krishnamurthy, and Vaish~\cite{ec/BarmanKV18} gave an $e^{1/e}+\epsilon\approx 1.445+\epsilon$ approximation.
This result is based on a notable link between the maximization of Nash social welfare (NSW) and EF1 (envy-free up to 1 item) allocations: Under identical valuations, any EF1 allocation achieves an approximation ratio of $e^{1/e}$ for the Nash social welfare objective. 
The authors developed an efficient algorithm that reduces the unweighted NSW problem with additive valuations to scenarios with identical valuations, and showed that the approximation ratio is at most this gap of $e^{1/e}$.
On the hardness side, Garg, Hoefer, and Mehlhorn showed that the problem is NP-hard to approximate within a ratio of $\sqrt{8/7}$~\cite{mor/Garg0M24}.

For weighted NSW with additive valuations, Brown, Laddha, Pittu, and Singh \cite{soda/BrownLPS24} developed a $5 \cdot \exp(2 \cdot D_{\text{KL}}\Big(w || \frac{\vec{1}}{n})\Big) = 5\cdot \exp(2\log n + 2\sum_{i \in N} w_i \log w_i)$ approximation algorithm, where $D_{\text{KL}}$ denotes the KL divergence of two distributions. In general, this ratio can be a super-constant. It was a longstanding open problem whether an $O(1)$-approximation exists. This was resolved in the affirmative by Feng and Li~\cite{icalp/FengLi24}, who gave an $(e^{1/e}+\epsilon)$-approximation for the problem, matching the best-known ratio for the unweighted case.
Their algorithm is based on a natural configuration LP relaxation for the problem and the Shmoys-Tardos rounding procedure, originally developed for the unrelated machine scheduling problem. 

For unweighted NSW with submodular valuations, Li and Vondrák~\cite{focs/LiV21} gave the first $O(1)$-approximation algorithm via convex programming. 
Subsequently, this ratio was improved to $(4 + \epsilon)$ by Garg, Husic, Li, Végh, and Vondrák~\cite{stoc/GargHLVV23}, using an elegant local search-based algorithm. On the hardness side, Garg, Kulkarni, and Kulkarni~\cite{talg/GargKK23} showed that the problem is NP-hard to approximate within $e/(e-1)$.
In the same paper, they also show that the $e/(e-1)$-approximation factor can be achieved when the number of agents is a constant.

For weighted NSW with submodular valuations, Garg, Husic, Li, Végh, and Vondrák~\cite{stoc/GargHLVV23} showed that their local search-based algorithm is $O(nw_{\max})$-approximate, where $w_{\max}$ is the maximum weight over the agents. 
It was open whether the problem admits an $O(1)$-approximation algorithm.
This was resolved by Feng, Hu, Li and Zhang~\cite{stoc/FHLZ25}, who presented a $(233+\epsilon)$-approximation algorithm.
Their algorithm is based on the configuration LP from~\cite{icalp/FengLi24}. They partition the fractionally assigned items into large and small items, and designed a randomized rounding procedure to ensure that the assignment of large items is a random matching, and the assignment of small items follows a pipage rounding procedure introduced by \cite{jco/AgeevS04}. 

\subsection{Our Results}

In this work, we make progress in improving the approximation ratio for Nash social welfare maximization with submodular valuations.
For weighted NSW with submodular valuations, we present a $(3.56+\epsilon)$ approximation algorithm (\cref{thm:ratio}) for an arbitrarily small $\epsilon>0$, which significantly improves the previous best-known ratio $(233+\epsilon)$ given by Feng, Hu, Li, and Zhang~\cite{stoc/FHLZ25}. It also simultaneously improves the approximation ratio of $(4+\epsilon)$ for unweighted NSW with submodular valuations by Garg, Husic, Li, Végh, and Vondrák~\cite{stoc/GargHLVV23}.

\begin{theorem}
For any $\epsilon>0$, there is a randomized $(3.56+\epsilon)$-approximation algorithm for the weighted Nash social welfare with submodular valuations, with running time polynomial $n^{\poly(1/\epsilon)}$.
\label{thm:ratio}
\end{theorem}

The approximation ratio of $(3.56+\epsilon)$ includes the losses from both solving the configuration LP and the rounding process. 
The upper bound of the integrality gap for the configuration LP considers only the rounding losses. 
In \cref{subsec:upper-integap}, we demonstrate that our rounding algorithm incurs a loss factor of just $3.45$, assuming the optimal fractional solution to the configuration LP is provided; this supports \cref{thm:integrality-gap}.
\begin{theorem}
The integrality gap of the configuration LP (\eqref{Conf-LP} in \cref{subsec:lp-solver}) is at most $3.45$.   
\label{thm:integrality-gap}
\end{theorem}

On the hardness side, we analyze the integrality gap of the natural configuration LP relaxation for the Nash social welfare problems, introduced by \cite{stoc/FHLZ25}. As this is the strongest known relaxation for these problems, understanding its limitations is crucial for the sake of algorithm design and algorithmic lower bounds.  
We show that the LP has an integrality gap at least $2^{\ln 2}-\epsilon \approx (1.617 - \epsilon)$ (\cref{thm:submodular_gap}) for weighted NSW with submodular valuations, which is slightly larger than the current best-known $e/(e-1) - \epsilon \approx (1.582 - \epsilon)$-hardness of approximation for the problem given by~\cite{talg/GargKK23}.
Our gap instance is built on a partition system proposed by~\cite{jacm/Feige98,algorithmica/KhotLMM08}, which is used to show the hardness for the submodular social welfare problem, whose goal is to maximize the sum of agents' utilities.

\begin{theorem}
For any constant $\delta > 0$, there is an instance of $\cI$ of weighted Nash social welfare with submodular functions, such that ${\fopt}/{\iopt}\geq 2^{\ln 2} - \delta$, where $\fopt$ is the exponential of the optimal value of the configuration LP (\eqref{Conf-LP} in \cref{subsec:lp-solver}) for $\cI$, and $\iopt$ is the optimal weighted Nash social welfare of $\cI$.  
Moreover, the valuations in $\cI$ are all coverage functions.
\label{thm:submodular_gap}
\end{theorem}

As a related result, we analyze the integrality gap of the configuration LP for \emph{additive} valuation functions. 
Using a similar construction, we prove that the gap is at least $(e^{1/e}-\epsilon)$ (\cref{thm:additive-gap}). 
This demonstrates that the $(e^{1/e}+\epsilon)$-approximation ratio given by~\cite{icalp/FengLi24} is tight, ruling out the possibility of obtaining a better approximation for the problem using this LP relaxation. 

\begin{theorem}
For any constant $\delta>0$, there is an instance $\cI$ of weighted Nash social welfare with additive functions, such that ${\fopt}/{\iopt}\geq e^{1/e}-\delta$, where $\fopt$ is the exponential of the optimal value of \eqref{Conf-LP} for $\cI$, and $\iopt$ is the optimal weighted Nash social welfare of $\cI$.
Moreover, the instance $\cI$ is a restricted assignment instance.
\label{thm:additive-gap}
\end{theorem}

Finally, for the unweighted NSW problem with additive valuations, we show a 4-agent instance with an integrality gap of $2^{1/4}-\epsilon\approx 1.189-\epsilon$. 
This is slightly larger than the current best-known $\sqrt{8/7}\approx 1.069$-hardness of approximation~\cite{mor/Garg0M24}.
We describe the instance in \cref{sec:unweighted-additive}. \medskip

We summarize our results in \cref{tab:submodular} and \cref{tab:additive}. In both tables, the integrality gap of \eqref{Conf-LP} for a problem is defined as the supremum of ${\fopt}/{\iopt}$ over all instances of the problem. 

\begin{table}[htb]
\centering
\begin{tabular}{cccc}
\hline
\multicolumn{4}{c}{Submodular Valuations}                                                                                   \\ \hline
\multicolumn{1}{c|}{}           & \multicolumn{1}{c|}{Hardness}        & \multicolumn{1}{c|}{Integrality Gap of \eqref{Conf-LP}} & Approximation Ratio      \\ \hline
\multicolumn{1}{c|}{Unweighted} & \multicolumn{1}{c|}{$e/(e-1)$~\cite{talg/GargKK23}} & \multicolumn{1}{c|}{}               & $3.56+\epsilon$~(\cref{thm:ratio}) \\ \hline
\multicolumn{1}{c|}{Weighted}   & \multicolumn{1}{c|}{}          & \multicolumn{1}{c|}{$\big[2^{\ln 2}, 3.45\big]$}    & $3.56+\epsilon$~(\cref{thm:ratio})  \\ \hline
\end{tabular}
\caption{Known results for Nash social welfare with submodular valuations, where $e/(e-1)\approx 1.582$, $2^{\ln 2}\approx 1.618$, and $\epsilon>0$ is an arbitrarily small constant. The $2^{\ln 2}$ lower bound on the integrality gap comes from \cref{thm:submodular_gap}. The $3.45$ upper bound comes from \cref{thm:integrality-gap}.
}
\label{tab:submodular}
\end{table}

\begin{table}[htb]
\centering
\begin{tabular}{cccc}
\hline
\multicolumn{4}{c}{Additive Valuations}                                                                                                       \\ \hline
\multicolumn{1}{c|}{}           & \multicolumn{1}{c|}{Hardness}           & \multicolumn{1}{c|}{Integrality Gap of \eqref{Conf-LP}}     & Approximation Ratio                 \\ \hline
\multicolumn{1}{c|}{Unweighted} & \multicolumn{1}{c|}{$\sqrt{8/7}-\epsilon$~\cite{mor/Garg0M24}} & \multicolumn{1}{c|}{[$2^{1/4}, e^{1/e}$] }                   & $e^{1/e}+\epsilon~\cite{ec/BarmanKV18}$          \\ \hline
\multicolumn{1}{c|}{Weighted}   & \multicolumn{1}{c|}{}             & \multicolumn{1}{c|}{[$e^{1/e}$,$e^{1/e}$]} & $e^{1/e}+\epsilon$~\cite{icalp/FengLi24} \\ \hline
\end{tabular}
\caption{Known results for Nash social welfare with additive valuations, where $\sqrt{8/7}\approx 1.069$, $e^{1/e}\approx 1.445$, $2^{1/4}\approx 1.189$, and $\epsilon>0$ is an arbitrarily small constant. The $2^{1/4}$ and $e^{1/e}$ lower bounds on the integrality gap for unweighted and weighted cases come from \cref{sec:unweighted-additive} and \cref{thm:additive-gap} respectively. The $e^{1/e}$ upper bound for both cases is due to the rounding algorithm of \cite{icalp/FengLi24}.
}
\label{tab:additive}
\end{table}

\subsection{Overview of Techniques}

In this section, we provide a high-level overview of the techniques used to obtain our main result, a $(3.56+\epsilon)$-approximation algorithm for the weighted Nash social welfare problem with submodular valuations. 

\paragraph{Solving Configuration LP with a Stronger Oracle.}
We use the same configuration LP as in \cite{stoc/FHLZ25}. As the separation oracle for the LP needs to solve the submodular maximization problem under a knapsack constraint, the LP solver loses a factor of $e/(e-1)$. 
To get a better approximation ratio, we design a stronger oracle with a refined approximation guarantee: we enumerate the $\poly(1/\epsilon)$ items with the largest values, defined using the greedy-order of items in the configuration, which repeatedly chooses the remaining item with the largest marginal gain. Thus, we only lose a factor of $e/(e-1)$ for the non-enumerated items (See \cref{lem:lp-solver} in \cref{subsec:lp-solver}). 
Roughly speaking, our approximation factor comes from three sources, though not in a simple multiplicative way: a loss of $e/(e-1)$ for solving the configuration LP, another $e/(e-1)$ factor from the multilinear extension, and an additional factor from applying concentration bounds. 
As is typical, the concentration bounds are strong when the marginal values of items are small. Thus, the third loss is incurred only for large items. With our new oracle, which loses the first factor only for small items, we can better align the first and third factors.

\paragraph{New Definition of Large Items.}
In the rounding algorithm of \cite{stoc/FHLZ25}, for each agent $i$, the largest item (the one with the largest $v_i(j)$) from each configuration $S$ is defined as a large item. 
The remaining fractional items assigned to $i$ are referred to as small items.  
Therefore, every agent $i$ gets precisely 1 fractional large item. 
In their randomized rounding algorithm, they ensure that the assignment of large items follows a random integral matching respecting the marginal probabilities, while small items are assigned via a pipage rounding procedure introduced in~\cite{jco/AgeevS04}.

In contrast, for each agent $i$, we choose the global 1 fractional largest item across all configurations (according to the marginal gains in the greedy order) as a large item, and the other fractional items as small items. 
So, the distribution of large items in configurations becomes less structured: some configurations may contain multiple large items, while others may contain none.  
On the other hand, our approach guarantees a useful monotonicity property: the size of any large item is at least that of any small item.  
In our analysis, we will reduce to the worst-case scenario where each configuration contains precisely one large item; therefore, the initial distribution of large items is not a major barrier. However, the monotonicity property allows us to give significantly stronger concentration bounds. 

\paragraph{Greedy Proxy Function.} As in \cite{stoc/FHLZ25}, we do a per-agent analysis and define two distributions of item sets for each agent $i$: the input distribution, which is the distribution of configurations from the configuration LP, and the output distribution, which is the distribution of the item sets assigned to $i$ by the algorithm.  
By making copies of items, we assume the item sets in the input distributions are disjoint. 
Another technical contribution we made is that we define a \emph{greedy proxy function} $\val_i$ (\cref{subsec:proxy-function}), which is a submodular function upper-bounded by the original valuation function $v_i$, with the following properties: $f_i$ is additive within each configuration, and $\val_i(j)$ for single item $j$ is its marginal gain in the greedy order for the configuration containing $j$. 
The two properties together imply that $\val_i(S) = v_i(S)$ for every configuration $S$. 
Therefore, replacing $v_i$ with $\val_i$ can only decrease the value of the output distribution, while maintaining that of the input distribution. 
The additivity property within configurations is useful for deriving stronger concentration bounds.

\paragraph{Analysis of Approximation Ratio.} In the analysis for the agent $i$, we normalize the valuation function $\val_i$ so that the smallest large item has value 1.  
Thus, the small items have a value of at most 1. Notice that in each configuration, there are $\poly(1/\epsilon)$ enumerated items (for simplicity, assume each configuration has at least $\poly(1/\epsilon)$ items). 
Therefore, for agent $i$, we have enumerated $\poly(1/\epsilon) \gg 1$ fractional items.  
In a typical situation, one would imagine that all the large items are enumerated, and the non-enumerated items are much smaller than 1. 
When this happens, our analysis gives an approximation ratio of $1 + (e/(e-1))^2 \approx 3.503$,  where 1 comes from the large items, and the first $e/(e-1)$ factor in the second term comes from solving the LP, and the other $e/(e - 1)$ factor comes from multilinear extension.  

Complications arise when some large items are not enumerated, or the sizes of some non-enumerated items are not sufficiently small. We were able to bound the approximation ratio for many cases by $1+(e/(e-1))^2$. 
The worst-case scenario is that there are a few small items of relatively large size (which we will refer to as medium items) that are not enumerated by our LP solution.
To obtain a tighter bound in this case, we need the help of the stronger concentration bound, which handles medium and small items together (\cref{lem:real-concentration} in \cref{subsec:real-concentration}).
The stronger concentration bound is a Chernoff-type left-tail bound under dependence, with the exponent separating the small (cap $\tau$) and medium portions, where $\tau$ is the minimum value of medium items. 
In effect, the small layer enjoys an exponentially strong concentration proportional to $\mu/\tau$, while the thin medium layer is charged separately.
The stronger concentration does not hold for general submodular functions; however, it is applicable in our case, thanks to the partial additivity property of the greedy proxy function.
Finally, we obtain the final approximation ratio of $3.56$, which is slightly larger than $1+(e/(e-1))^2$.

We formulate the task of analyzing the difference between the expected logarithmic values of the input and output distributions as a mathematical program (MP). Through a sequence of worst-case analyses, we gradually simplify the MP to many linear programs (LPs) with continuous variables. By discretizing these variables, we obtain LPs with discrete variables, which we then solve computationally. This yields our approximation ratio of $3.56 + \epsilon$.

\section{Preliminaries}
\label{sec:preliminaries}

\begin{definition}[Monotone Submodular Function]

Let $M$ be the item set and $f:2^{M}\to\R_{\geq 0}$ be a function defined over $M$.
\begin{itemize}
    \item $f$ is monotone if $f(S)\leq f(T)$ whenever $S\subseteq T$.
    \item $f$ is submodular if for any $S,T\subseteq M$, $f(S)+f(T) \geq f(S\cap T) + f(S\cup T)$.
\end{itemize}
\end{definition}

For convenience, we slightly abuse the notation and use $f(j)$ to represent $f(\set{j})$.
In this work, we shall frequently work on the {\em greedy order} of elements (\cref{def:greedy-order}).

\begin{definition}[Greedy Order]
Given an item set $S\subseteq M$ and a monotone submodular function $f:2^M\to\R_{\geq 0}$, the greedy order $(e_1,\ldots,e_{\abs{S}})$ of items in $S$ under $f$ is defined as: the item at the $t$-th position is the one that maximizes the marginal increment (breaking tie arbitrarily), i.e., 
\[
e_t \gets \argmax_{e\in S} f(\set{e_1,\ldots,e_{t-1}}\cup\set{e}) - f(\set{e_1,\ldots,e_{t-1}}).
\]
For an item $e_t$, we call the value of $f(\set{e_1,\ldots,e_{t-1}}\cup\set{e_t}) - f(\set{e_1,\ldots,e_{t-1}})$ as the greedy-marginal-gain under this greedy order.
\label{def:greedy-order}
\end{definition}

In this work, we shall frequently use the following two continuous extensions of a submodular function: multilinear and concave extensions of a submodular function, which are widely used in the literature.
The concave extension is also known as the ``Configuration LP'' extension in the literature.
\begin{definition}[Multilinear Extension]
Given a monotone submodular function $f:2^M\to \R_{\geq 0}$, its multilinear extension $F:[0,1]^m \to \R_{\geq 0}$ is defined as:
\[
F(\bx) := \sum_{S\subseteq M} f(S) \prod_{j\in S} x_j \prod_{j\in (M\setminus S)}(1-x_j).
\]
\end{definition}

\begin{definition}[Concave Extension]
Given a monotone submodular function $f:2^M\to\R_{\geq 0}$, its concave extension $f^+:[0,1]^m\to\R_{\geq 0}$ is:
\[
f^+(\bx) := \max\left\{ \sum_{S\subseteq M} \alpha_S f(S): \sum_{S\subseteq M} \alpha_S \leq 1;\forall S\subseteq M, \alpha_S\geq 0; \forall j\in M, \sum_{S:j\in S} \alpha_S = x_j  \right\}.
\]
\label{def:concave}
\end{definition}
The following relation between the multilinear and concave extensions has been shown by Vondrák~\cite[Lemma 3.7, 3.8]{vondrakthesis}.
\begin{lemma}[\cite{vondrakthesis}]
Given a monotone non-negative submodular function $f$, its multilinear extension $F$, and its concave extension $f^+$, and a point $\bx\in[0,1]^m$, we have $f^+(\bx) \geq F(\bx) \geq (1-\frac{1}{e})\cdot f^+(\bx)$.   
\label{lem:concave-multilinear}
\end{lemma}

Given an arbitrary vector $\bx\in[0,1]^m$, the value of $F(\bx)$ is also equal to the expected value of the random set $X$ produced by a {\em product distribution}, where each item $j$ in $X$ is sampled independently with probability $x_j$.

\subsection{Greedy Proxy Function} 
\label{subsec:proxy-function}

In the analysis, we shall use an auxiliary function for the valuation function, which is called {\em Greedy Proxy Function}.
This auxiliary function is only used in the analysis and shall provide a lower bound of the solution returned by the algorithm while maintaining the value of the LP.

Let $v:2^{U}\to\R_{\geq 0}$ be a monotone submodular function defined over an element set $U$ with $\abs{U}:=n$.
Let $P_1,\ldots,P_k$ be a partition of $U$.
For each element set $P_i$, let $\pi^i$ be the greedy order of elements in $P_i$. 
Let $\pi^i(t)\in P_i$ be the element at the $t$-th position and $\pi^i([t])\subseteq P_i$ be the first $t$ elements ($\pi^i([0])=\emptyset$).

\begin{lemma}
There exists a set function $\val:2^U\to\R_{\geq 0}$ such that $\val$ has the following properties:
\begin{enumerate}[leftmargin=*,label=(\ref{lem:greedy_proxy}\ablue{\alph*})]
    \item $\val$ is a non-negative monotone submodular function.
    \label{prop:monotone_submodular}
    \item $\val(S)\leq v(S)$ for all $S\subseteq U$; 
    \label{prop:lower_bound}
    \item $\val(\pi^i(t))=v(\pi^i([t]))-v(\pi^i([t-1]))$ for all $1 \leq t \leq |P_i|$ and $i\in[k]$;
    \label{prop:marginal_value}
    \item Fix an arbitrary partition $P_i$, for each $S\subseteq P_i$, we have $\val(S)=\sum_{j\in S}\val(j)$.
    \label{prop:additive}
\end{enumerate}
\label{lem:greedy_proxy}
\end{lemma}

We prove in \cref{app:greedy-proxy} that the function $\val$ defined in \cref{def:greedy_proxy} satisfies all the properties outlined in \cref{lem:greedy_proxy}.
\begin{definition}
For each element $j\in U$, assuming $j\in P_i$ and $j$ is at the $t$-th position on $\pi^i$, we define $\phi_{j} := v(\pi^i([t]))-v(\pi^i([t-1]))$.
Now, we define $\val$ as follows:
\[
\val(S):=\min_{T\subseteq S}\left\{ v(T)+\sum_{j\in S\setminus T} \phi_j \right\}, \quad \forall S\subseteq U.
\]
\label{def:greedy_proxy}
\end{definition}

We remark that the proof of \cref{lem:greedy_proxy} works for arbitrary orderings $\pi^i, i \in [k]$ of sets $P_i,i \in [k]$. However, we shall apply the lemma, more precisely \ref{prop:marginal_value}, for the greedy orders.
More details can be found in \cref{app:greedy-proxy}.
In the analysis, we shall replace the agent's valuation function with \cref{def:greedy_proxy} as \ref{prop:additive} is a crucial property to our analysis.
Additionally, \ref{prop:lower_bound} indicates that $\val$ serves as a lower bound for the algorithm's solution, while \ref{prop:marginal_value} ensures that the value of the LP's solution remains unchanged.

\subsection{Concentration Bound}
\label{subsec:concentration}

Feng, Hu, Li, and Zhang prove that the random variables produced by the following modified pipage rounding (\cref{alg:pipage}) satisfy one of the relaxed left tails of the Chernoff bound~\cite[Theorem 2.7]{stoc/FHLZ25}.
In this work, we shall use other bounds on exponential functions (\cref{thm:pipage-rounding}), which are more convenient for us to perform operations later, and are able to lead to a better approximation ratio.
Similar to~\cite {stoc/FHLZ25,soda/HarveyO14}, the proof is mainly based on the {\em concave pessimistic estimator} technique but still needs some other tools from~\cite{focs/ChekuriVZ10}.

\begin{algorithm}[h]
    \caption{Modified Pipage Rounding~\cite{stoc/FHLZ25}}
    \begin{algorithmic}[1]
        \Require{$\bx^* \in [0, 1]^{\bar n}$.}
        \Ensure{an integral $\bx \in \{0, 1\}^{\bar n}$.}
        \State $\bx \gets \bx^*$.
        \While{$\bx$ is not integral}
            \State do one of the two operations arbitrarily:
            \Statex\hspace*{\algorithmicindent}{\bf Operation 1:}
            \State\hspace*{\algorithmicindent}choose one coordinate $a \in [\bar n]$ with $x_a \in (0, 1)$, two reals $\delta_1 \in (0, x_a]$ and $\delta_2 \in (0, 1- x_a]$.
            \State\hspace*{\algorithmicindent}with probability $\frac{\delta_2}{\delta_1 + \delta_2}$ \textbf{do}: $x_a \gets x_a - \delta_1$,  \textbf{else do}: $x_a \gets x_a + \delta_2$.
            \Statex\hspace*{\algorithmicindent}{\bf Operation 2:}
            \State\hspace*{\algorithmicindent}choose two distinct coordinates $a, b \in [\bar n]$ with $x_a, x_b \in (0, 1)$, two reals $\delta_1 \in (0, \min\{x_a, 1 - x_b\}]$, and $\delta_2 \in (0, \min\{1 - x_a, x_b\}]$.
            \State\hspace*{\algorithmicindent}with probability $\frac{\delta_2}{\delta_1 + \delta_2}$ \textbf{do}: $x_a \gets x_a - \delta_1, x_b \gets x_b + \delta_1$.
            \State\hspace*{\algorithmicindent}\hspace*{75pt} \textbf{else do}: $x_a \gets x_a + \delta_2, x_b \gets x_b - \delta_2$.
      \EndWhile
      \State \Return $\bx$
    \end{algorithmic}
        \label{alg:pipage}
\end{algorithm}

\begin{lemma}
Consider \cref{alg:pipage} and assume it always terminates in finite number of iterations. Let $\bx^{*}\in[0,1]^{\bar n}$ and $\bx\in\set{0,1}^{\bar n}$ be its input and output, then $\E[f(\bx)]\geq F(\bx^*)$.
\label{lem:pipage+convex}
\end{lemma}

The correctness of \cref{lem:pipage+convex} is mainly due to the function $F(\cdot)$ being linear on the direction $\be_a$ and convex on the direction $(\be_a-\be_b)$ for any $a,b\in[m]$, which was proved by~\cite[Page 6]{siamcomp/CalinescuCPV11}.
Proofs analogous to that of \cref{lem:pipage+convex} can be found in~\cite[Lemma 3.5]{siamcomp/CalinescuCPV11} and~\cite[Lemma A.2]{stoc/FHLZ25}.

\begin{theorem}
    \label{thm:pipage-rounding}
    Let $v:2^{[\bar n]} \to \R_{\geq 0}$ be a monotone submodular function with marginal values in $[0,1]$, and $F:[0, 1]^{\bar n} \to \R_{\geq 0}$ be the multilinear extension of $v$. 
    Fix any $\tau\in(0,1)$, let $S:=\set{i\in[\bar n]\mid v(i)\leq \tau}$ and $L:=[\bar n]\setminus S$ be two item sets (a partition of $[\bar n]$).
    
    Let $\bx\in [0, 1]^{\bar n}$ and $\bx^\smm$ be the vector $\bx$ that is restricted to the item set $S$. Let $\mu := F(\bx)$, $\mu^\smm:=F(\bx^\smm)$.
    Suppose that $\bx^\inte \in \{0, 1\}^{\bar n}$ be the output of \cref{alg:pipage} for the input $\bx$, and $U = \{i \in [\bar n]: x_i^\inte = 1\}$.
    Then, for any $\lambda < 0$, we have 
    \begin{align*}
        \E\left[ e^{\lambda \cdot v(U)} \right] \leq e^{(e^{\lambda\cdot\tau}-1)\cdot\mu^\smm/\tau+(e^{\lambda}-1)(\mu-\mu^\smm)}.
    \end{align*}
\end{theorem}

We remark that the concentration bound in \cref{thm:pipage-rounding} does not hold when $\lambda>0$. 
However, if either $U$ is obtained through independent rounding, i.e., an element $i$ is included in $U$ independently with a probability $x_i$, or $v$ is an additive function, then the concentration bound holds for any $\lambda \in \mathbb{R}$. More details can be found in \cref{app:concentration}.

\section{The Algorithm}
\label{sec:alg}

Throughout the paper, we shall let ${{\epsilon}} > 0$ be a fixed sufficiently small constant with $1/{{\epsilon}}$ being an integer. Our goal is to obtain a $3.56 + O({{\epsilon}})$ approximation for the problem, which can be turned into a $3.56 + {{\epsilon}}$ approximation by scaling down ${{\epsilon}}$.
We define $\epsilon_1:=\epsilon^{20}$.

\subsection{Solving Configuration LP}
\label{subsec:lp-solver}

The weighted NSW with submodular valuations admits a natural configuration LP \eqref{Conf-LP}, which was first introduced by~\cite{stoc/FHLZ25,icalp/FengLi24}.
In the LP, we have a variable $y_{i,S}\in\set{0,1}$ for every agent $i\in N$ and an item set $S\subseteq M$, indicating whether the set of items assigned to $i$ is precisely $S$. 
The objective is to maximize the logarithm of the weighted Nash social welfare.
The first constraint ensures that each item is assigned to exactly one agent.
The second constraint ensures that each agent gets exactly one item set.
\begin{align}
    \text{max} && \sum_{i\in N,S\subseteq M} w_i \cdot  y_{i,S} \cdot \ln(v_i(S)) & \tag{\text{Conf-LP}} \label{Conf-LP}\\
    \text{s.t.} \nonumber \\
    &&\sum_{S:j\in S}\sum_{i\in N}y_{i,S} &= 1, &\forall j\in M\label{LPC:item} \\
    &&\sum_{S\subseteq M}y_{i,S} &= 1, &\forall i\in N \label{LPC:agent}\\
    &&y_{i,S} &\geq 0, &\forall i\in N, S\subseteq M 
\end{align}

Feng, Hu, Li and Zhang~\cite{stoc/FHLZ25} gave a separation oracle for the dual of \eqref{Conf-LP}, based on the $e/(e-1)$-approximation algorithm for the submodular maximization with a knapsack constraint problem due to \cite{orl/Sviridenko04}.
Namely, they proved that \eqref{Conf-LP} can be solved in polynomial time within an additive error of $\ln (e/(e-1)+{\epsilon_1})$~\cite[Lemma 3.1]{stoc/FHLZ25}. 

In this work, we shall present another stronger separation oracle that gives us a solution with more structural insights:
We only need to lose the factor of $e/(e-1)$ on small items, while the large items lose nothing.
Such a structural fractional solution enables us to present a better rounding algorithm.

For each item set $S\subseteq M$, let $\pi^i(S)$ be the greedy order of items in $S$ under the valuation function $v_i$.
Let $S_i^{\enu}$ be the first $\min\{\abs{S},1/{\epsilon_1}\}$ items on the greedy order $\pi^i(S)$. (``$\enu$'' stands for ``enumerate'' as in \cref{def:enu-nonenu}.) The proof of \cref{lem:lp-solver} is deferred to \cref{app:lp-solver}.

\begin{lemma}
Let $\mathsf{LP}$ be the optimal value of \eqref{Conf-LP}.
For any constant ${\epsilon_1}>0$, there exists a polynomial time algorithm that finds a fractional solution $\by^*:=(y^*_{i,S})_{i\in N,S\subseteq M}$ such that  
\[
\sum_{i,S}w_i \cdot y^*_{i,S} \cdot \ln \left( v_i(S^{\enu}_i) + \frac{e}{e-1}\cdot \left( v_i(S)-v_i(S^{\enu}_i) \right) \right) \geq \mathsf{LP}-\ln(1-{\epsilon_1}).
\]
\label{lem:lp-solver}
\end{lemma}

To achieve \cref{lem:lp-solver}, we need to use the modified greedy algorithm for the submodular knapsack problem~\cite{orl/Sviridenko04} as a separation oracle, which shall enumerate all sets of items up to a size $1/{\epsilon_1}$ (instead of $3$).
This requires $n^{O(1/{\epsilon_1})}=n^{O(1/{{\epsilon}}^{20})}$ time.

\subsection{Bipartite Multigraph and Rounding}
\label{subsec:rounding}

Our rounding algorithm is also based on the construction of a bipartite multigraph $G:=(N\cup M, E)$, which is similar to~\cite{stoc/FHLZ25}, but with a modification of some key components.
After solving \eqref{Conf-LP} using \cref{lem:lp-solver}, we obtain a solution $\by^*:=(y^*_{i,S})_{i\in N, S\subseteq M}$ which only consists of non-zero coordinates.
Note that we only have a polynomial number of non-zero coordinates.

In~\cite{stoc/FHLZ25}, for every $i, S$ with $y^*_{i, S} > 0$, they choose the item $j$ with the largest $v_i(j)$. 
They then create a marked edge of value $y^*_{i, S}$ between $i$ and $j$, and unmarked edges of value $y^*_{i, S}$ between $i$ and all items in $S \setminus \set{j}$. 
Parallel marked (unmarked) edges are merged. Therefore, there could be at most two parallel edges between an agent $i$ and an item $j$, one marked and one unmarked. 

We construct the bipartite multigraph $G$ in a different way.  
For every agent $i$, instead of choosing one largest item from each configuration for marked edges, we choose the overall 1 fractional largest items from all configurations, where the value of an item in a configuration is its marginal gain in the greedy order. 
So, an item $j$ may have many different values in different configurations for the same agent $i$.

We describe the bipartite multigraph $G$ in a formally; each edge in $G$ is either \emph{marked} or \emph{unmarked}. 
Till the end of \cref{subsec:rounding}, we fix an agent $i \in N$ and show how to construct the edges between $i$ and $M$ in $G$.  
We sort all pairs $\{(S, j): y^*_{i,S}> 0, j \in S\}$ non-decreasing order of $\phi^S_j$ values, where $\phi_j^S$ is item $j$'s greedy marginal gain in $S$. 
Let $(S_1, j_1), (S_2, j_2), \ldots, (S_q, j_q)$ be the order. Let $z_t := y^*_{i, S_t}$ for every $t \in [q]$. 
Let $t^*$ be the smallest index such that $z([t^*]) \geq 1$.  
Let $a := 1 - z([t^*-1]) > 0$ and $b := z([t^*])-1 \geq 0$; so $a + b = z_{t^*}$. 

\begin{figure}[htb]
    \centering
    \includegraphics[width=1\linewidth]{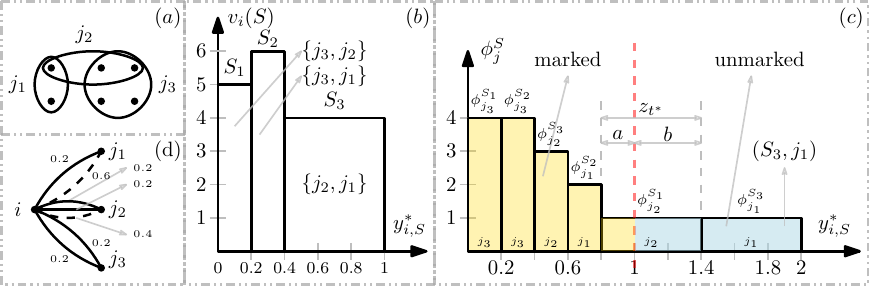}
    \caption{Illustration for the constructed bipartite multigraph. The figure only shows one agent $i$. We assume that one agent's valuation function is a coverage function, as shown in the subfigure (a), where each item is a set and the valuation of an item set is equal to the size of the union of these items. The subfigure (b) shows a fractional solution to \eqref{Conf-LP}. Each rectangle represents a configuration, where the height is $v_i(S)$ and the width is $y^*_{i,S}$. Inside each configuration, items have been sorted in greedy order. The subfigure (c) shows how to build the bipartite multigraph. Each rectangle stands for a pair $(S,j)$, where the height is $\phi_{j}^{S}$ and the width is $y^*_{i,S}$; so every rectangle would correspond to one or two edges in the bipartite graph. The yellow and blue rectangles then correspond to marked and unmarked edges, respectively. The subfigure (d) shows the constructed bipartite multigraph, where solid and dashed edges are marked and unmarked edges, respectively.}
    \label{fig:large-small-items}
\end{figure}

\begin{enumerate}[label=(\arabic*)]
    \item For each $t < t^*$ with $S_t \neq S_{t^*}$ we create a marked edge between $i$ and $j_t$ of $x$-value $z_t = y^*_{i, S_t}$.
    \item For each $t < t^*$ with $S_t = S_{t^*}$ we create a marked edge between $i$ and $j_t$ of $x$-value $a$; if $b > 0$ we create another marked edge between $i$ and $j_t$ of $x$ value $b$. 
    \item For each $t > t^*$ with $S_t \neq S_{t^*}$, we create an unmarked edge between $i$ and $j_t$ of $x$-value $z_t = y^*_{i, S_t}$.
    \item For each $t > t^*$ with $S_t = S_{t^*}$, we create an unmarked edge between $i$ and $j_t$ of $x$-value $a$; if $b > 0$ we create another unmarked edge between $i$ and $j_t$ of $x$-value $b$.
    \item Consider $t = t^*$. We create a marked edge between $i$ and $j_t$ of $x$-value $a$. If $b  > 0$, we create a unmarked edge between $i$ and $j_t$ of $x$-value $b$.
\end{enumerate}
We remark that for (2) (resp.\ (4)), we can simply create a marked (resp.\ unmarked) of weight $a+b = z_{t^*}$. Then the case will be the same as case (1) (resp.\ (3)). This will not affect the algorithm. However, later in the analysis, we need to make two copies of the configuration $S_{t^*}$, one with fraction $a$ and the other with fraction $b$. It is convenient to split the edges for all items in $S_{t^*}$, not just $j_{t^*}$. Unlike the construction in \cite{stoc/FHLZ25}, we do \emph{not} merge parallel marked (unmarked) edges.  An example is shown in \cref{fig:large-small-items}.

We then apply the rounding algorithm proposed by~\cite[Algorithm 2]{stoc/FHLZ25} to the constructed bipartite multigraph. 
In each iteration, the rounding algorithm shall find either a marked cycle or a pseudo-marked path: a marked cycle is a simple cycle of marked edges, and a pseudo-marked path is a simple path of marked edges between two items $j$ and $j'$, with one unmarked edges $ij$ concatenated at the beginning, and another unmarked edge $j'i'$ concentrated at the end. 
Then it randomly rotates the cycle or shifts the path. 
It maintains the marginal probabilities, and the properties that every agent is incident to 1 fractional marked edge, and every item is incident to 1 fractional edge. 
The rounding algorithm terminates with an integral solution in polynomial time. 
We let $\intx \in \set{0,1}^{E}$ be the random integral solution returned by the algorithm.

\section{Per-Client Analysis of Algorithm}
\label{sec:analysis}

In this section, we analyze the approximation ratio achieved by the algorithm.
For every agent $i \in N$, let $\vlp^i$ be the unweighted contribution of agent $i$ to the LP solution, that is,
\[
\vlp^i:=\sum_{S\subseteq M}y^*_{i,S}\cdot\ln\left( v_i(S^{\enu})+\frac{e}{e-1}\cdot\left( v_i(S)-v_i(S^{\enu}) \right) \right).
\]
Recall that $G:=(N\cup M, E)$ is the constructed bipartite multigraph with an edge value vector $\bx\in[0,1]^E$, and $\intx \in \set{0,1}^E$ is the random integral solution returned by the rounding algorithm.
Let $T_i:=\set{j \in M: \exists ij \in E, \intx_{ij}=1}$ be the set of items assigned to agent $i$ in the solution $\intx$. 
Let $\valg^i$ be the expected logarithmic value the agent $i$ obtained:
\[
\valg^i:=\E[\ln v_i(T_i)].
\]

The remainder of \cref{sec:analysis} and \cref{subsec:comparsion} aims to prove the following claim:
\begin{lemma}
For any agent $i\in N$, we have
$\valg^i \geq \vlp^i - \ln \left( 3.56+{{\epsilon}} \right).$
\label{lem:ratio}
\end{lemma}

We now prove \cref{thm:ratio} based on \cref{lem:ratio}.

\begin{proofof}{\cref{thm:ratio}}
The following proof is the same as the one in~\cite[Theorem 1.1]{stoc/FHLZ25}.
By \cref{lem:ratio} and the linearity of expectation and $\sum_{i\in N}w_i=1$, we have:
\begin{align*}
\E\left[\ln v_i(T_i) \right] &= \valg^i \geq \vlp^i - \ln \left( 3.56+{{\epsilon}} \right) \\
&= \sum_{S\subseteq M}y^*_{i,S}\cdot\ln\left( v_i(S^{\enu})+\frac{e}{e-1}\cdot\left( v_i(S)-\val(S^{\enu}) \right) \right) - \ln\left( 3.56+{{\epsilon}} \right).
\end{align*}
Taking the exponential of both sides and applying Jensen's inequality, we have:
\begin{align*}
    &\quad \E\left[ \prod_{i\in N}v_i(T_i)^{w_i} \right]\\
    &\geq \exp\left( - \ln(3.56+{{\epsilon}}) \right) \cdot \exp\left( \sum_{S\subseteq M}y^*_{i,S}\cdot\ln\left( v_i(S^{\enu})+\frac{e}{e-1}\cdot\left( v_i(S)-v_i(S^{\enu}) \right) \right) \right)\\
    &\geq \frac{1-{\epsilon_1}}{3.56 + {{\epsilon}}} \cdot \opt = \frac{1}{3.56 + O({{\epsilon}})} \cdot \opt.
\end{align*}
where the last inequality is by \cref{lem:lp-solver} and ${{\epsilon}}={\epsilon_1}^{1/20}$.
The running time is polynomial in $1/{{\epsilon}}^{20}$ because \cref{lem:lp-solver} requires a modified greedy algorithm for the submodular knapsack problem.
\end{proofof}

Till the end of the \cref{subsec:comparsion}, we fix an agent $i$. We aim to prove \cref{lem:ratio} for $i$, by comparing $\vlp^i$ with $\valg^i$ for a fixed agent $i\in N$. 
In \cref{subsec:assumption}, we show that we can, without loss of generality, make assumptions on the LP solution from \cref{lem:lp-solver} and on the valuation functions that simplify analysis without affecting the algorithm. 
We view the fractional solution as a distribution with mean $\vlp^i$, and the rounding procedure as inducing another distribution; they are formally defined in \cref{subsec:input-output}. 
We upper bound the input distribution in \cref{subsec:input-dist} and use a natural lower bound for the output distribution, then prove in \cref{subsec:comparsion} that their gap is only $(\ln 3.56+{{\epsilon}})$, completing \cref{lem:ratio}. 
The proof relies on a sequence of mathematical programs and computer-assisted checks; see \cref{subsec:comparsion}.

\subsection{Making Copies of Items and $v_i$ Additive within Configurations}
\label{subsec:assumption}
As the same item $j$ will have different greedy-marginal-gain in different configurations, it will be convenient for us to treat the same item in different configurations for $i$ as different copies.  This will guarantee that the configurations for $i$ are disjoint. 

Recall that in the construction of the incident edges of $i$ in $G$ in \cref{subsec:rounding}, we have an order $(S_1, j_1), (S_2, j_2), \ldots, (S_q, j_q)$ of pairs, $z_t = y^*_{i, S_t}$ for every $t \in [q]$, $t^*$ is smallest integer with $z([t^*]) \geq 1$, $a = 1 - z([t^*-1])$ and $b = z([t^*])-1$.

We take $\cS$ as the set of all configurations $S$ with $y^*_{i,S}>0$, and we say that the $y^*$-value of an $S\in\cS$ is $y^*_{i,S}$. If $b > 0$, we apply the following additional step: we split $S_{t^*}$ into two copies of itself, with $y^*$-values $a$ and $b$, respectively. In this case, $\cS$ is a multiset.

We treat the same item $j$ in different sets $S\in\cS$ as different copies of $j$; let $M'$ be the set of all copies of items appearing in $\cS$. Therefore, $\cS$ forms a partition of $M'$. Moreover, there is a one-to-one correspondence between $M'$ and the incident edges of $i$ in the graph $G$. 
The value of the function $v_i$ can be naturally extended to copies of items denoted by $v_i'$: the value of a set $S'$ (e.g. $v'(S')$) of copies is the same as $v_i(S)$, where $S$ is the set of items that have at least one copy in $S'$.

It is easy to see that the extended valuation function $v_i'$ is still monotone and submodular. When $j$ is assigned to $i$ by our algorithm, we can see which edge $ij$ has $\intx_{ij} = 1$, and thus the copy of $j$ assigned to $i$. As two copies of the same item do not appear in the same configuration for $i$, neither will they be assigned to $i$ simultaneously by our algorithm, the extension of $v_i$ to the set of copies has no effect on the value of $\vlp^i$ and $\valg^i$.  We also remark that making copies is feasible only because we are focusing on a fixed agent $i$.

With configurations forming a partition of $M'$ disjoint, we can now make another assumption.  
We can pretend that the valuation function $v_i$ is indeed its greedy proxy function $\val_i$ with respect to the partition $\cS$ (\cref{subsec:proxy-function}).
The assumption maintains the value of $\vlp^i$ due to the properties \ref{prop:marginal_value} and \ref{prop:additive} of the greedy proxy function. 
It does not increase the value of $\valg^i$ because $\val_i(S)\leq v_i(S)$ for any item set $S\subseteq M'$ \ref{prop:lower_bound}. 
A useful property about $\val_i$ is that it is additive within each configuration $S$ \ref{prop:additive}.

\paragraph{Notations.}
Since we shall focus on a fixed agent $i \in N$ till the end of \cref{sec:MPs}, we omit the subscripts $i$ from most of the notations from now on.  We shall use $\valg$ and $\vlp$ 
to denote $\valg^i$, $\vlp^i$. Abusing notations slightly, we shall use $M$ for $M'$, as we do not need to use the global set of items anymore.  Recall that each $S \in \cS$ has a $y^*$-value, and now we use $y^*_S$ to denote it. This is obtained from the original $(y^*_{i, S})_{S}$ by possibly splitting one configuration into two copies, and making copies of items. 
$T \subseteq M$ will be the random set of items assigned to $i$ by the algorithm.  
$\val$ is the greedy proxy function that was used to replace the original function $v_i$.
Let $\bx \in [0, 1]^{M}$ be the vector $\bx$ restricted to the incident edges of $i$ in $G$, which has a one-to-one correspondence with $M$.
We emphasize the following two properties:
\refstepcounter{theorem}
\label{property:assumption}
\begin{enumerate}[leftmargin=*,label=(\ref{property:assumption}\ablue{\alph*})]
    \item The sets in $\cS$ form a partition of $M$.  We have $y^*_S > 0$ for every $S \in \cS$ and $\sum_{S \in \cS} y^*_S = 1$.
    \item $\val$ is additive inside each $S\in\cS$, i.e., $\val(R)=\sum_{j\in R}v_j$ for any $R\subseteq S$ with $S\in\cS$. 
\end{enumerate}

\subsection{Input and Output Distributions}
\label{subsec:input-output}

The rounding algorithm and LP solution naturally define two partitions of the item set $M$, which are shown in \cref{def:large-small} and \cref{def:enu-nonenu}, respectively.
We define two partitions of $M$ as follows:

\begin{definition}[Large, Medium, Small Items]
We say $j \in M$ is a large item if its corresponding edge $ij$ in $G$ is marked, and non-large otherwise. 
Let $M^{\lg}$ and $M^\nlg$ denote the sets of large and non-large items, respectively. 
$M^\nlg$ is further partitioned into medium and small items: an item $j\in M^{\nlg}$ is a medium item if $\val(j)\geq\sqrt{{\epsilon_1}}$ and a small item if $\val(j)<\sqrt{{\epsilon_1}}$.
Let $M^{\md}$ and $M^{\sm}$ be the medium and small items.
\label{def:large-small}    
\end{definition}

For each $S\in\cS$, recall that $S^{\enu}$ is the first $\min\{\abs{S},1/{\epsilon_1}\}$ items in non-increasing order of $\val(j)$ values, as we replaced the original valuation function for $i$ with its proxy function, which we now denote by $\val$.

\begin{definition}[Enumerated and Non-enumerated Items]\label{def:enu-nonenu}
For each $S \in \cS$, we say items in $S^\enu$ are enumerated. Let $S^\nenu := S \setminus S^\enu$ and call items in $S^\nenu$ non-enumerated.
\end{definition}

By scaling the valuation functions, we assume $\min_{j \in M^{\lg}} \val(j) = 1$.
This implies $\max_{j \in M^{\nlg}} \val(j) \leq 1$. 
$M^{\lg}$, $M^{\nlg}$ in \cref{def:large-small} correspond to marked and unmarked edges incident to $i$ in $G$ respectively.

\begin{figure}[htb]
    \centering
    \includegraphics[width=0.85\linewidth]{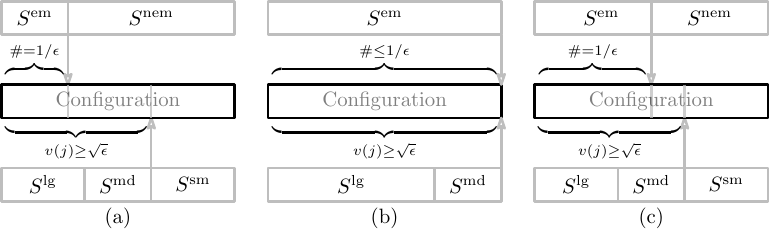}
    \caption{Illustration of the configuration type defined by the LP and rounding partitions, where ``\#'' represents the number of items. It is possible that some sets of $S^{\enu}$, $S^{\nenu}$, $S^{\lg} $, $S^{\md}$, $S^{\sm}$ are empty, e.g., as shown in subfigure (b), $S^{\nenu}$ and $S^{\sm}$ are empty.}
    \label{fig:two-partitions}
\end{figure}

For each configuration $S$, let $S^{\lg}:=S\cap M^{\lg}$, $S^{\md}:=S\cap M^{\md}$ and $S^{\sm}:=S\cap M^{\sm}$. Then, for each configuration $S \in \cS$, the two partitions $(S^{\lg}, S^\md, S^\sm)$ and $(S^\enu, S^\nenu)$ are consistent with the greedy order of $\val$-values of $S$.  
There are many configuration types depending on the relation between the two partitions, some of which are shown in \cref{fig:two-partitions}.

\paragraph{Output Distribution.}
Recall that $T\subseteq M$ is the set of items assigned to agent $i$ by the algorithm.
Let $T^{\lg}:=M^{\lg}\cap T$ and $T^{\nlg}:=M^{\nlg}\cap T$.
The rounding algorithm naturally defines a probability distribution on $T^{\lg}$ and $T^{\nlg}$.
In the analysis, we shall use the following lower bound of the output distribution:
\begin{equation}
\valg=\E[\ln (\val(T^{\lg}\cup T^{\nlg}))] \geq \E\left[\ln\left(\max\left\{  \val(T^{\lg}) , \val(T^{\nlg}) \right\} \right)\right].
\label{equ:lower-bound-output}
\end{equation}

\paragraph{Input Distribution.}
The input distribution is defined in a straightforward way: choose $S$ randomly with probability $y^*_S$.
Note $\val$ is additive within each configuration.
Thus, we have
\begin{equation}
\vlp=\E\left[\ln\left( \val(S^{\enu})+\frac{e}{e-1}\cdot\left( \val(S)-\val(S^{\enu}) \right) \right)\right]=\E\left[\ln\left( \val(S^{\enu})+\frac{e}{e-1}\cdot\val(S^{\nenu}) \right)\right].
\label{equ:input-dist}
\end{equation}

Let $\bx^{\nlg},\bx^{\sm},\bx^{\md}$ be the vector $\bx \in [0, 1]^M$ that is restricted to the items in $M^{\nlg},M^{\sm},M^{\md}$, respectively. Define 
\begin{align*}
\mu^{\sm}:= F(\bx^{\sm}) \qquad \text{and } \qquad \mu^{\nlg}:=F(\bx^{\nlg}),
\end{align*}
where $F$ is the multilinear extension of the function $\val$. 
For each configuration $S$, let $S^{\nlg}:=S\cap M^{\nlg}$.
As proved by~\cite{stoc/FHLZ25}, the returned solution $T^{\lg},T^{\nlg}$ satisfies the properties stated in \cref{lem:rounding-properties}.

\begin{lemma}[\cite{stoc/FHLZ25}]
The random sets $T^{\lg}$ and $T^{\nlg}$ satisfy the following properties:
\begin{enumerate}[leftmargin=*,label=(\ref{lem:rounding-properties}\ablue{\alph*})]
        \item $\Pr[|T^{\lg}| = 1] = 1$. 
        \label{property:large-item}
        \item For every $j \in M$, we have $\Pr[j \in T] = x_{j}$. 
        \label{property:marginal-prob}
        \item $\E\left[\val(T^{\nlg})\right]\geq F(\bx^{\nlg})\geq (1-1/e)f^+(\bx^{\nlg})\geq (1-1/e)\E[\val(S^{{\nlg}})]$, where $f^+$ is the concave extension of $\val$.  
        \label{property:rounding-lp}
        \item \label{property:concentration-thm} The following inequality holds for every $\lambda <0$: 
        \[
        \E\Big[e^{\lambda \cdot \val(T^{\nlg})}\Big]\leq e^{(e^\lambda-1)(\mu^{\nlg}-\mu^{\sm})+(e^{\lambda\sqrt{{\epsilon_1}}-1})\mu^{\sm}/\sqrt{{\epsilon_1}}}.
        \]
    \end{enumerate}
\label{lem:rounding-properties}
\end{lemma}

We note that \ref{property:large-item} and \ref{property:marginal-prob} are the direct consequences of the rounding algorithm in~\cite{stoc/FHLZ25}.
The property \ref{property:concentration-thm} follows from~\cref{thm:pipage-rounding} by noting $\val(j)\leq\sqrt{{\epsilon_1}}$ for all $j\in M^{\sm}$.
The first inequality in \ref{property:rounding-lp} is due to \cref{lem:pipage+convex}, i.e., the modified pipage rounding also loses nothing compared to the multilinear extension.
The second inequality in \ref{property:rounding-lp} is due to \cref{lem:concave-multilinear}, and the last inequality in \ref{property:rounding-lp} follows from the definition of concave extension; a proof can be found in the proof of Lemma 4.7 in~\cite{stoc/FHLZ25}.

\subsection{Separating Small and Large $p^*$ Cases}
\label{subsec:input-dist}

We define $p^*$ as the probability of the event that there exist non-small items in the configuration $S$ that are not enumerated by the LP solver:
\begin{align*}
    p^*:=\sum_{S:S^{\sm}\subsetneq S^{\nenu}}y^*_S.    
\end{align*}

When $p^*$ is small, we establish the following upper bound on $\vlp$:
\begin{lemma}
If $p^*\leq {\epsilon_1}^{1/3}$, then we have
\[
\vlp \leq \E\left[ \ln\left(\val(S^{\lg})+\val(S^{\md})+\frac{e}{e-1}\cdot \val(S^{\sm}) \right)\right] +O({\epsilon_1}^{1/3}).
\]
\label{lem:upper-bound-input-small}
\end{lemma}

\begin{proof}
    We compare $\ln(\val(S^{\enu}) + {e}/{(e-1)} \cdot \val(S^\nenu))$ and $\ln(\val(S^{\lg}) + \val(S^\md) + {e}/{(e-1)} \cdot \val(S^\sm))$. The former is at most the latter if $S^\nenu \subseteq S^\sm$.  Otherwise, we have $S^\sm \subsetneq S^\nenu$, and the former is at most the latter plus $\ln {e}/{(e-1)} = \Theta(1)$. The lemma follows from that $\E[S^\sm \subsetneq S^\nenu] = p^* \leq {\epsilon_1}^{1/3}$. 
\end{proof}

On the other hand, if $p^*$ is large, we can prove that the expected value of small and medium items is large (\cref{lem:upper-bound-input-large}). This implies that small and medium items are almost perfectly concentrated, in which we can exploit the concentration property and prove that $\vlp$ and $\valg$ have a small gap. 
Specifically, when $p^*$ is large, it is not necessary to further distinguish the medium and small items; one can just imagine that all $f(S^\md)$ in \cref{lem:upper-bound-input-small} are scaled by $e/(e-1)$.
We still need to set up a mathematics program to bridge the gap between the relaxed bounds for $\vlp$ and $\valg$, but we do not need the assistance of a computer program.
We can directly bound the gap of $\vlp$ and $\valg$ based on \cref{lem:upper-bound-input-large}; see \cref{subsec:large-p*} for details.

\begin{lemma}
If $p^*>{\epsilon_1}^{1/3}$, then we have $\E[\val(S^{\md})] = \Omega(1/{\epsilon_1}^{1/6})$.   
\label{lem:upper-bound-input-large}
\end{lemma}

\begin{proof}Notice that
    $\E[|S^{\lg} \cup S^{\md}|] \geq \sum_{S: S^\sm \subsetneq S^\nenu} y^*_S \cdot 1/{\epsilon_1} = p^* \cdot 1/{\epsilon_1} \geq {\epsilon_1}^{1/3} \cdot 1/{\epsilon_1} = {\epsilon_1}^{-2/3}.$
    On the other hand, we have $\E[|S^{\lg}|] = 1$. So, 
    $\E[|S^{\md}|] \geq {\epsilon_1}^{-2/3} - 1$. Every $j \in S^{\md}$ has $\val(j) \geq \sqrt{{\epsilon_1}}$. By the additive property of the greedy proxy function, we have
    $\E[\val(S^{\md})] \geq ({\epsilon_1}^{-2/3} - 1)\sqrt{{\epsilon_1}} = {\epsilon_1}^{-1/6} - \sqrt{{\epsilon_1}} = \Omega({\epsilon_1}^{-1/6})$.
\end{proof}

\subsection{Deriving Concentration Bounds}
\label{subsec:real-concentration}
For convenience, we define $t^{\nlg}:=\val(T^{\nlg})$, $t^{\md}:=\val(T^{\md})$ and $t^{\sm}:=\val(T^{\sm})$.
Note that the valuation function $\val$ is not necessarily additive inside $T$ and thus we can only have $t^{\nlg}\leq t^{\md}+t^{\sm}$ by the submodularity.
We shall use the concentration bound \ref{property:concentration-thm}, which has a term of $\mu^{\nlg}-\mu^{\sm}$.
This makes us unable to apply the concentration bound directly because $\mu^{\nlg}-\mu^{\sm}\ne\mu^{\md}$ as $\val$ is not additive.
To fix this issue, we need to work on a slightly weaker concentration bound (\cref{lem:real-concentration}).

To this end, we define 
\begin{align*}
    \mubar^{\md}:=(1-1/e)\E[\val(S^{\md})], \quad
    \mubar^{\sm}:=(1-1/e)\E[\val(S^{\sm})],\quad\text{and}\quad
    \mubar^{\nlg}:=(1-1/e)\E[\val(S^{\nlg})].
\end{align*}
Since $\val$ is additive on $S$ (i.e. \ref{prop:additive}), we have $\mubar^{\nlg}=\mubar^{\md}+\mubar^{\sm}$ by the linearity of the expectation.
Note that $\mu^{\sm}\geq\mubar^{\sm}$, $\mu^{\md}\geq\mubar^{\md}$ and $\mu^{\nlg}\geq\mubar^{\nlg}$ by \ref{property:rounding-lp}.
In the following, we first prove \cref{lem:real-concentration-key}, which enables us to prove \cref{lem:real-concentration}.
The bound stated in \cref{lem:real-concentration} is the concentration that we will use in our analysis.

\begin{lemma}
$\E[t^{\nlg}]\geq \mu^{\nlg} \geq \mubar^{\sm}+\mubar^{\md}$.    
\label{lem:real-concentration-key}
\end{lemma}

\begin{proof}
The lemma is a direct consequence of \ref{property:rounding-lp} and the additivity property of the valuation function:
\[
\E[t^{\nlg}]=\E\left[\val(T^{\nlg})\right]\geq F(\bx^{\nlg})=\mu^{\nlg}\geq (1-1/e)f^+(\bx^{\nlg})\geq (1-1/e)\E[\val(S^{{\nlg}})]=\mubar^{\sm}+\mubar^{\md}.
\]
\end{proof}

\begin{lemma}
 For any $\lambda<0$, we have 
$\E\left[e^{\lambda t^{\nlg}}\right]\leq e^{(e^{\lambda}-1)\mubar^{\md}+(e^{\lambda\sqrt{{\epsilon_1}}}-1)\mubar^{\sm}/\sqrt{{\epsilon_1}}}$.
\label{lem:real-concentration}
\end{lemma}
\begin{proof}
By \ref{property:concentration-thm}, for any $\lambda<0$, we have
\begin{align*}
\E\left[e^{\lambda t^{\nlg}}\right]
&=\E\Big[e^{\lambda \cdot \val(T^{\nlg})}\Big]
\leq \exp\left((e^\lambda-1)(\mu^{\nlg}-\mu^{\sm})+(e^{\lambda\sqrt{{\epsilon_1}}-1})\mu^{\sm}/\sqrt{{\epsilon_1}}\right) \tag*{\ref{property:concentration-thm}}   \\
&= \exp\left( \underbrace{\left(\frac{e^{\lambda \sqrt{{\epsilon_1}}}}{\sqrt{{\epsilon_1}}}-(e^{\lambda}-1)\right)}_{\leq 0}\cdot\mu^{\sm} + \underbrace{\left(e^{\lambda}-1\right)}_{\leq 0}\cdot\mu^{\nlg} \right)  \\
&\leq \exp\left( \left(\frac{e^{\lambda \sqrt{{\epsilon_1}}}}{\sqrt{{\epsilon_1}}}-(e^{\lambda}-1)\right)\cdot\mubar^{\sm} + \left(e^{\lambda}-1\right)\cdot\mu^{\nlg} \right)\tag{$\mu^{\sm}\geq\mubar^{\sm}$} \\
&\leq\exp\left( \left(\frac{e^{\lambda \sqrt{{\epsilon_1}}}}{\sqrt{{\epsilon_1}}}-(e^{\lambda}-1)\right)\cdot\mubar^{\sm} + \left(e^{\lambda}-1\right)\cdot\left(\mubar^{\sm}+\mubar^{\md}\right) \right)\tag{\cref{lem:real-concentration-key}} \\
&=\exp\left((e^\lambda-1)\mubar^{\md}+(e^{\lambda\sqrt{{\epsilon_1}}-1})\mubar^{\sm}/\sqrt{{\epsilon_1}}\right).
\end{align*}
\end{proof}

\section{Analysis of Gap Using Mathematical Programs} \label{sec:MPs}
\label{subsec:comparsion}

In \cref{subsec:input-dist}, we separate two cases by the value of $p^*$.
When $p^*\leq {\epsilon_1}^{1/3}$,  we aim to upper bound the following difference:
\[
\E\left[ \ln\left(\val(S^{\lg})+\val(S^{\md})+\frac{e}{e-1}\cdot \val(S^{\sm}) \right)\right] - \E\left[\ln\left(\max\left\{  \val(T^{\lg}), \val(T^{\nlg})\right\}\right)\right].
\]
While we will upper bound the following difference if $p^*>{\epsilon_1}^{1/3}$:
\[
\E\left[ \ln\left(\val(S^{\lg})+\frac{e}{e-1}\cdot \val(S^{\nlg}) \right)\right] - \E\left[\ln\left(\max\left\{  \val(T^{\lg}), \val(T^{\nlg})\right\}\right)\right],
\]
in which we no longer need to distinguish medium and small items.
Combining these two cases gives an upper bound of $\vlp-\valg$ up to an additive error of $O({{\epsilon}})$, which finishes the proof of \cref{lem:ratio}; see \cref{subsec:proofofratio} for details.

To bound the difference in the small $p^*$ case, we need \mpref{mp:one} while we need the other simpler mathematical program (\mpref{mp:four}) for the large $p^*$ case.
In the following, we shall focus on the small $p^*$ case and aim to derive an upper bound on \mpref{mp:one}, while we will bound the large $p^*$ in \cref{subsec:large-p*}.
All analyses presented in \cref{subsec:unique-t}-\ref{subsec:waterfilling} also work for the large $p^*$ case; see \cref{subsec:large-p*} for details.

In \cref{subsec:upper-integap}, we will upper bound the integrality gap of our rounding algorithm, and thus, we assume that we are given an optimal fractional solution, where we no longer need to lose $1-1/e$ for non-large items.
In this case, we aim to compare the difference between:
\[
\E\left[ \ln\left(\val(S^{\lg})+ \val(S^{\nlg}) \right)\right] - \E\left[\ln\left(\max\left\{  \val(T^{\lg}), \val(T^{\nlg})\right\}\right)\right],
\]
where we need a different but even simpler mathematics program (\mpref{mp:five}), but the analysis in \cref{subsec:unique-t}-\ref{subsec:waterfilling} still works; see \cref{subsec:upper-integap} for details.

Now, we focus on the small $p^*$ case till the end of \cref{subsec:computer-program}.
From our analysis in \cref{sec:analysis}, the difference of the small $p^*$ case is bounded by the value of the following mathematical program:
\mppara{1}{mp:one}{
Given a partition $M^{\lg},M^{\md},M^{\sm}$ of $M$, a fixed sufficiently small constant ${\epsilon_1}>0$ and a set of configurations $\cS$, we aim to maximize
\begin{equation}
\E\left[\ln\left(\val(S^{\lg})+\val(S^{\md})+\frac{e}{e-1}\cdot \val(S^{\sm})\right)\right]-\E\left[\ln\left(\max\left\{  \val(T^{\lg}) , t^{\nlg} \right\} \right)\right],
\label{equ:ratio-1}
\end{equation}
subject to the following constraints:
\begin{enumerate}[label=(MP1\alph*),leftmargin=*]
    \item $S^{\lg},T^{\lg}\subseteq M^{\lg}$, $S^{\md}\subseteq M^{\md}$, $S^{\sm}\subseteq M^{\sm}$ are random sets and $t^{\nlg}\in\R_{\geq 0}$ is a random variable.
    \label{mp1:1}
    \item $\Pr[|T^{\lg}|=1]=1$.
    \label{mp1:large-item-size}
    \item For any item $j$, $\Pr[j\in S^{\lg}]=\Pr[j\in T^{\lg}]$.
    \label{mp1:large-item-marginal}
    \item $\val:2^M\to\R_{\geq 0}$ is a monotone submodular function and $\val$ is additive inside each configuration $S\in\cS$.
    \label{mp1:additive}
    \item $\forall j\in M^{\lg}$, $\val(j)\ge 1$; $\forall j\in M^{\md}, \val(j)\in[\sqrt{{\epsilon_1}},1)$; $\forall j\in M^{\sm}$, $\val(j)\in[0,\sqrt{{\epsilon_1}})$.
    \label{mp1:large-item-value}
    \item $\E[t^{\nlg}]\ge \mubar^{\md}+\mubar^{\sm}$, where $\mubar^{\md}:=(1-1/e)\E[\val(S^{\md})],\mubar^{\sm}:=(1-1/e)\E[\val(S^{\sm})]$.
    \label{mp1:mu}
    \item For any $\lambda<0$, $\E\left[e^{\lambda t^{\nlg}}\right]\leq e^{(e^{\lambda}-1)\mubar^{\md}+(e^{\lambda\sqrt{{\epsilon_1}}}-1)\mubar^{\sm}/\sqrt{{\epsilon_1}}}$.
    \label{mp1:concentration}
\end{enumerate}}
From now on, we analyze \mpref{mp:one}. 

\subsection{Overview of Analysis}
Our goal is to upper bound the gap between the relaxed input distribution and the output distribution produced by rounding, phrased as the \mpref{mp:one}. 
We progressively transform both distributions to worst-case surrogates that only increase the objective of \mpref{mp:one}, until the problem depends on only a few parameters.
At that point, the remaining bound is verified by a computer program. 
The major property that is required by these reductions is the concavity of the logarithmic function.
The high-level worst-case reductions include the following three main steps:

\paragraph{(1) Align $T^{\lg}$ with the non-large value $t^{\nlg}$ (\cref{subsec:align:1},\ref{subsec:align:2}).}
This step aims to obtain the property of $\val(T^{\lg})=t^{\nlg}\geq 1$ so that we can drop the parameter of $T^{\lg}$.
Using the structural property from \cref{subsec:unique-t}, which ties each large outcome to a unique non-large mass, we first locally edit the distribution in \cref{subsec:align:1} so that the value of the large part always dominates the non-large value at level 1, i.e., ensure $\val(T^{\nlg})=\max\{1,t^{\nlg}\}$.
Namely, whenever we have $\val(T^{\lg})<\max\{1,t^{\nlg}\}$, we shall modify the value of the item in $T^{\lg}$.
This will not break the constraints, but would increase the objective because of the monotonicity and concavity of the log function.

We then can remove the case where $t^{\nlg}<v(T^{\lg})=1$ by shifting the probability mass between events with $t^{\nlg}<1$ and $t^{\nlg}>1$ in \cref{subsec:align:2}.
So, we finally will have either $t^{\nlg}<1$ or $t^{\nlg}>1$ with probability 1.
The former case is easy to bound because the output value would be fixed, and the input value can be easily bounded based on $t^{\nlg}\leq 1$ and \ref{mp1:mu}.

\paragraph{(2) Match the laws of $S^{\lg}$ and $T^{\lg}$ (\cref{subsec:s-t-large-same}).}
This step aims to show that $S^{\lg}$ and $T^{\lg}$ can be regarded as the same distribution so that we can drop the parameter of $S^{\lg}$.
We first rebalance the probability mass between configurations with $|S^{\lg}|\ge 2$ and $|S^{\lg}|=0$ so that $|S^{\lg}|=1$ with probability 1. 
By the concavity of the log function, the movement from multi-large to empty does not decrease the input term, hence the LP draws $S^{\lg}$ and the rounded outcome $T^{\lg}$ have the same distribution by \ref{mp1:large-item-marginal}.

\paragraph{(3) Water-filling the non-large items (\cref{subsec:waterfilling}).}
This step aims to replace $\val(S^{\md})+(e/(e-1))\val(S^{\sm})$ with a single parameter.
Let $s_1:=\val(S^{\lg})$ and $s_2:=\E[\val(S^{\md})+(e/(e-1))\val(S^{\sm})\mid S^{\lg}]$.
Among all feasible reallocations of the non-large mass across configurations with a fixed total expected value, the worst case for the input term $\E[\ln(s_1+s_2)]$ is attained by the water-filling manner: $s_2^*(t)=(k-s_1(t))^{+}$.
This follows from the concavity of the log function: a more balanced $(s_1,s_2)$ pair would lead to a larger log value.

\begin{figure}
    \centering
    \includegraphics[width=1\linewidth]{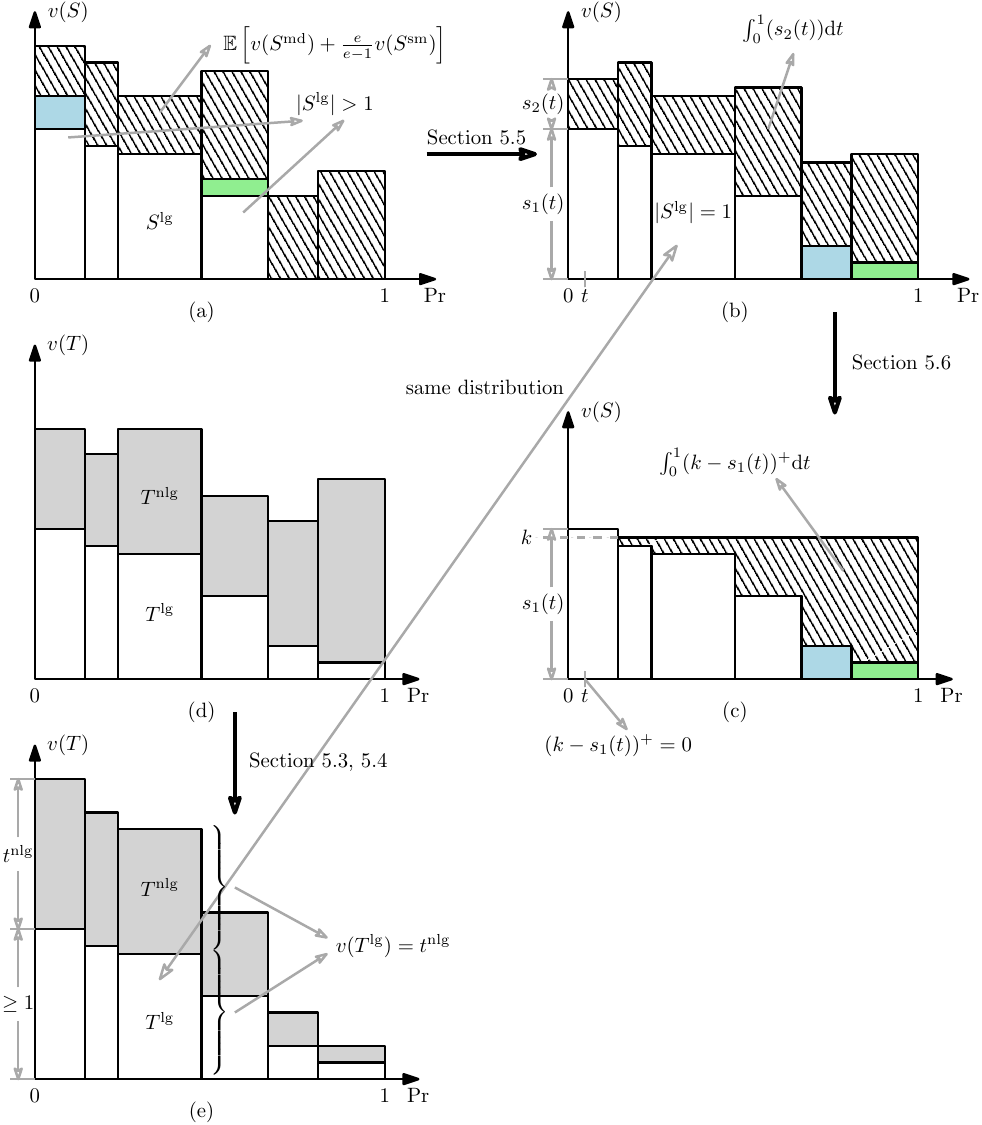}
    \caption{Overview of Analysis of \mpref{mp:one}. The Figures (a), (b), (c) are input distributions, and (d), (e) are output distributions. After a series of worst-case reductions, the input and output distributions finally become (c) and (e), respectively. In \cref{subsec:computer-program}, we compare (c) and (e) via a computer program.}
    \label{fig:overview}
\end{figure}

The above worst-case transformation is shown in \cref{fig:overview}.
After these three steps, \mpref{mp:one} will be transformed into \mpref{mp:two}, which depends on only a few parameters; see \cref{subsec:mp2} for details.

\subsection{Making Copies of Large Items}
\label{subsec:unique-t}

In the following way, we can add the following property to \mpref{mp:one}.
\begin{enumerate}[leftmargin=*, label=(MP1h)]
    \item For every $j \in M^{\lg}$, there is a unique $t$ with $\Pr[T^{\lg} = j, t^{\nlg} = t] > 0$.
    \label{mp1:unique-t}
\end{enumerate}

Suppose that the property does not hold: There exists $j \in M^{\lg}$ and two different values $t$ and $t'$ with $\Pr[T^{\lg} = \{j\}, t^\nlg = t] > 0$ and $\Pr[T^{\lg} = \{j\}, t^\nlg = t'] > 0$. 
Then, we add a copy $j'$ of $j$ with $\val(j') = \val(j)$ to $M^{\lg}$. 
We replace the event $T^{\lg} = \{j\} \wedge t^\nlg = t'$ with $T^{\lg} = \{j'\} \wedge t^\nlg = t'$. 
We can easily modify the input distribution so that \ref{mp1:large-item-marginal} holds, without changing the value of \mpref{mp:one}.

\subsection{Guaranteeing $\val(T^{\lg})=\max\{t^{\nlg},1\}$}
\label{subsec:align:1}

We show that we can, without loss of generality, assume that $\Pr[\val(T^{\lg}) = \max\{t^{\nlg}, 1\}] = 1$ in \mpref{mp:one}. 
Note that $\val(T^{\lg})\geq 1$ must hold due to \ref{mp1:large-item-value}.
For the sake of contradiction, suppose that for some $j \in M^{\lg}$ and $t \in \R_{\geq 0}$ with $\val(j) > \bar t := \max\{t, 1\}$, we have $\Pr[T^{\lg} = \{j\}, t^\nlg = t] > 0$. 

For convenience in notation, let $\Gamma:= \val(S^{\md})+e/(e-1)\cdot \val(S^{\sm})$.
Then consider the following operation: decrease $\val(j)$ to $\bar t$. 
This will not violate the properties \ref{mp1:1}-\ref{mp1:concentration}. 
It decreases $\E\left[\ln\left(\val(S^{\lg})+\Gamma\right)\right]$ in \eqref{equ:ratio-1} by
\[
\Pr\left[j \in S^{\lg}\right] \cdot \E\left[\ln\left(\val(S^{\lg}) + \Gamma\right) - \ln\left(\val(S^{\lg} \setminus j) + \bar t + \Gamma\right) \Big| j \in S^{\lg}\right]
\leq \Pr\left[j \in S^{\lg}\right] \cdot (\ln \val(j) - \ln \bar t),
\]
where $\val(j)$ denotes its old value and the inequality used $\ln(a + c) - \ln(b + c) < \ln b - \ln c$ for every $a > b > 0$ and $c > 0$. 
The operation decreases $\E\left[\ln \max\left\{\val(T^{\lg}), t^{\nlg}\right\}\right]$ in \eqref{equ:ratio-1} by 
\[
\quad\Pr\left[T^{\lg} = \{j\}, t^{\nlg} = t\right] \cdot \left(\ln \max\left\{\val(j), t\right\} - \ln\max\left\{\bar t, t\right\}\right)
= \Pr\left[T^{\lg} = \set{j}\right] \cdot (\ln \val(j) - \ln \bar t),
\]
due to property \ref{mp1:unique-t} and $\val(j) > \bar t \geq t$.  
As $\Pr[j \in S^{\lg}] = \Pr[j \in T^{\lg}] = \Pr[T^{\lg} = \set{j}]$ by \ref{mp1:large-item-marginal} and \ref{mp1:large-item-size}, the decrement to the positive term of \eqref{equ:ratio-1} is at most the decrement to the negative term. 
So, the operation can only increase \eqref{equ:ratio-1}.

Then consider the case where $\Pr[t^{\lg} = \{j\}, t^{\nlg} = t] > 0$ for some $j \in M^{\lg}$ and $t > \val(j)$. 
We then consider the operation of increasing $\val(j)$ to $t$. 
Again, this will not affect the properties \ref{mp1:1}-\ref{mp1:concentration}.  
It can only increase $\E\left[\ln\big(\val(S^{\lg})+\Gamma\big)\right]$ in \eqref{equ:ratio-1}. 
However, it does not change the term $\E\left[\ln \max\big\{\val(T^{\lg}), t^{\nlg}\big\}\right]$ as $t > \val(j)$. So, the operation can only increase \eqref{equ:ratio-1}.

So, we can apply the above two operations repeatedly until $\Pr[t^{\lg} > \max\{t^{\nlg}, 1\}] = \Pr[t^{\lg} < t^{\nlg}] = 0$. 
So, we always have $t^{\nlg} < t^{\lg} = 1$ or $1 \leq t^{\lg} = t^{\nlg}$, which is equivalent to $t^{\lg} = \max\{t^{\nlg}, 1\}$.

\subsection{Guaranteeing $\val(T^{\lg})=t^{\nlg}\geq 1$}
\label{subsec:align:2}

In \cref{subsec:align:1}, we already show that $\val(T^{\lg})$ is equal to $\max\{t^{\nlg},1\}$ in the worst case.
In this section, we show that we can further, without loss of generality, assume $\val(T^{\lg}) = t^{\nlg} \geq 1$ in \mpref{mp:one} by removing the possibility of $t^{\nlg} < \val(T^{\lg}) = 1$.  

First, we assume $\Pr[\val(T^{\lg}) = 1, \val(t^{\nlg}) = t] > 0$ for at most one value $t \in [0, 1]$. 
Otherwise, let $t_{\mathrm{av}} = \E[t^{\nlg}|\val(T^{\lg}) = 1, t^{\nlg} \leq 1]$. For every $j \in M^{\lg}, \val(j) = 1$ and $t \leq 1$ with positive $\Pr[T^{\lg} = \{j\}, t^{\nlg} = t]$, we move the probability mass to the event $\Pr[T^{\lg} = \{j\}, t^{\nlg} = t_{\mathrm{av}}]$. 
This does not change the objective \eqref{equ:ratio-1}, and it does not violate the properties in \mpref{mp:one}. 
It does not violate the property $\val(T^{\lg}) = \max\{t^{\nlg}, 1\}$ established in \cref{subsec:align:1}.  
In particular, this operation does not change the input distribution, so $\bar\mu^{\sm}$ and $\bar\mu^{\md}$ remain the same.
Hence, the concentration \ref{mp1:concentration} still holds as the left side of the inequality only decreases. 

Assume $p := \Pr[\val(T^{\lg}) = 1, t^{\nlg} = t] > 0$ for some unique $t \in [0, 1)$, and $p' = \Pr[T^{\lg} = \{j\}, t^{\nlg} = t'] > 0$ for some $j \in M^{\lg}$ and $t'$ with $\val(j) = t' > 1$. 
Let $a > 0$ be the largest number such that $t + qa \leq 1$ and $t' - pa \geq 1$.  
We then move the probability mass of any event $T^{\lg} = \{j'\} \wedge t^{\nlg} = t$ with $\val(j') = 1$ to the event $T^{\lg} = \{j'\} \wedge t^{\nlg} = t + qa$, and the probability mass of the event $T^{\lg} = \{j\} \wedge t^{\nlg} = t'$ to the event $T^{\lg} = \{j\} \wedge t^{\nlg} = t'-pa$. 
This does not change the value of \eqref{equ:ratio-1}, or break the properties of \mpref{mp:one}. 
Again, as before, the input distribution does not change, and the concentration bound in \ref{mp1:concentration} still works. 
The operation may break the property established in the last step $\val(T^{\lg}) = \max\{t^{\nlg}, 1\}$, but we can apply the operation in the last step again to hold the property.

Therefore, repeatedly applying the operation if possible, we have either $\Pr[t^{\nlg} > 1] = 0$ or $\Pr[t^{\nlg} < 1] = 0$. 
In the former case, we have $\Pr[\val(T^{\lg}) = 1, t^{\nlg} = t] = 1$ for some $t \leq 1$, in which we claim that \eqref{equ:ratio-1} can be bounded by $\ln(1+(e/(e-1))^2)$.
\begin{lemma}
If $\Pr[\val(T^{\lg}) = 1, t^{\nlg} = t] = 1$ with $t\leq 1$, then the value of \eqref{equ:ratio-1} is at most $\ln(1+(e/(e-1))^2)$.
\label{lem:algin}
\end{lemma}
\begin{proof}
Note that under the condition of the lemma, we have $\ln\left(\max\{\val(T^{\lg}),t^{\nlg}\}\right)=\ln 1=0$ with probability $1$.
Thus, \eqref{equ:ratio-1} is at most: 
\begin{align*}
\eqref{equ:ratio-1}&=\E\left[\ln\left(\val(S^{\lg})+\val(S^{\md})+\frac{e}{e-1}\cdot \val(S^{\sm})\right)\right] \leq \E\left[\ln\left(\val(S^{\lg})+\frac{e}{e-1}\cdot\left( \val(S^{\md})+\val(S^{\sm})\right)\right)\right]  \\
&=\E\left[\ln\left(\val(S^{\lg})+\left(\frac{e}{e-1}\right)^2\cdot\left(\bar\mu^{\md}+\bar\mu^{\sm}\right)\right)\right] \tag{the definition of $\bar\mu^{\md},\bar\mu^{\sm}$} \\
&\leq \E\left[\ln\left(\val(S^{\lg})+\left(\frac{e}{e-1}\right)^2\cdot \E[t^{\nlg}]\right) \right] \tag{\ref{mp1:mu}} \\
&\leq \E\left[\ln\left(\val(S^{\lg})+\left(\frac{e}{e-1}\right)^2\right)\right] \tag{$t^{\nlg}\leq 1$ with probability 1} \\
&\leq \ln\left( \E\left[\val(S^{\lg})\right]+\left(\frac{e}{e-1}\right)^2\right) \tag{Jensen's inequality} \\
&=\ln\left(1+\left(\frac{e}{e-1}\right)^2\right) \tag{\ref{mp1:large-item-marginal}, \ref{mp1:additive} and $\val(T^{\lg})=1$ with probability 1}
\end{align*}
This proves the lemma.
\end{proof}

With \cref{lem:algin}, it remains to focus on the latter case, where we always have $\val(T^{\lg}) = t^{\nlg} \geq 1$.

\subsection{Guaranteeing $\abs{S^{\lg}}=1$}
\label{subsec:s-t-large-same}

We show that assuming $\abs{S^{\lg}}=1$ with probability 1 does not decrease the value of \eqref{equ:ratio-1} nor impact the constraints in \mpref{mp:one}.
Suppose that there is a configuration $S_a\in\cS$ with $\abs{S_a^{\lg}}\geq 2$.
Then, there must exist another configuration $S_b\in\cS$ with $\abs{S_b^{\lg}}=0$.
This is due to $\sum_{S\in\cS}y^*_S=1$ (LP constraint \eqref{LPC:agent}), and we have only one unit of large items in $M^{\lg}$.
If $\abs{S^{\lg}}\geq 1,\forall S\in\cS$ and $\exists \bar S\in\cS$ with $\abs{\bar S^{\lg}}\geq 2$, we shall have more than one unit of large items.

For notation convenience, let $s^{\lg}:=\val(S^{\lg})$, $s^{\jt}:=\val(S^{\md})+(e/(e-1))\val(S^{\sm})$ and $s:=\val(S)$.
Now, let $S$ and $\bar S$ be the configurations such that $\abs{S^{\lg}}\geq 2$ and $\abs{\bar S^{\lg}}=0$. 
Let $\beta$ be the minimum probability among configurations $S$ and $\bar S$, i.e., $\beta:=\min\{y^*_S,y^*_{\bar S}\}$.
Consider an arbitrary item $j\in S$, we distinguish two cases, note that $\beta>0$ and $S\setminus\set{j}\ne\emptyset$.

\paragraph{Case \Rom{1}:} $\val(j)+s^{\jt}>\bar s^{\jt}$. We decrease the probability of $S$ and $\bar S$ by $\beta$ and increase the probability of the following configurations by $\beta$: $\set{j}\cup S^{\md}\cup S^{\sm}$ and $(S^{\lg}\setminus \set{j})\cup \bar S^{\md}\cup \bar S^{\sm}$. This will not affect the properties in \mpref{mp:one} but it would increase \eqref{equ:ratio-1} because:
\[
\beta\cdot \left(\ln(\underbrace{\val(j)+s^{\jt}}_{\text{(\rom{1})}})+\ln(\underbrace{\val(S\setminus\set{j})+\bar s^{\jt}}_{\text{(\rom{2})}})\right)> \beta\cdot\left(\ln(\underbrace{\val(S)+s^{\jt}}_{\text{(\rom{3})}})+\ln(\underbrace{\bar s^{\jt}}_{\text{(\rom{4})}})\right).  
\]

This inequality holds because (a) $\beta>0$; (b) $(\rom{1})+(\rom{2})=(\rom{3})+(\rom{4})$; (c) $(\rom{2}), (\rom{1})>(\rom{4})$ and $(\rom{1}), (\rom{2})<(\rom{3})$; (d) the function $\ln(\cdot)$ is concave, so a more balanced pair $((\rom{1}),(\rom{2}))$ has a higher value compared to a more imbalanced pair $((\rom{3}),(\rom{4}))$.

\paragraph{Case \Rom{2}:} $\val(j)+s^{\jt}\leq \bar s^{\jt}$.
In this case, we decrease the probability of $S$ and $\bar S$ by $\beta$ and increase the probability of the following configurations by $\beta$: $(S^{\lg}\setminus\set{j}, \val(j)+s^{\jt})$ and $(\set{j},\bar s^{\jt}-\val(j))$.
This will not affect the properties of \mpref{mp:one}, and the value of \eqref{equ:ratio-1} does not change.
Note that the two new configurations above may not be able to correspond to the ``physical'' configuration, but they maintain the value of \eqref{equ:ratio-1}.
One can just imagine that we move the item $j$ from $S$ to $\bar S$ and then artificially increase and decrease the value of $s^{\jt}$ and $\bar s^{\jt}$ by $\val(j)$, respectively.

Repeated application of the operation ensures $\Pr[\abs{S^{\lg}}]=1$, without affecting the properties in \mpref{mp:one} or increasing the value of \eqref{equ:ratio-1}.
Notice that by \ref{mp1:large-item-marginal}, the distributions of $S^{\lg}$ and $T^{\lg}$ then will be identical.

\subsection{Water Filling Upper Bound}
\label{subsec:waterfilling}

We use the notation $(z)^{+}$ ($z\in\R$) to represent $\max\{z,0\}$.
Consider the input distribution (e.g., \cref{fig:overview}(b)), let $t\in[0,1]$ be an arbitrary point in the $x$-axis.
Note that the order of configurations does not matter.
Let $s_1(t)$ and $s_2(t)$ be value of $\val(S^{\lg})$ and $\val(S^{\md})+(e/(e-1))S^{\sm}$, respectively, where $S$ is the configuration contains the point $t$.
In this notation, we have 
\[
\E\left[\ln\left(\val(S^{\lg})+\val(S^{\md})+\left(\frac{e}{e-1}\right)\cdot \val(S^{\sm})\right)\right]=\int_{0}^{1}\ln (s_1(t)+s_2(t))\rmd t.
\]
\begin{lemma}
There exists some $k\geq 0$ with $\int_{0}^{1}(k-s_1(t))^+\rmd t=\E[\val(S^{\md})+(e/(e-1))\val(S^{\sm})]$ such that $\int_{0}^{1}\ln(s_1(t)+s_2(t))\rmd t \leq \int_{0}^{1}\ln(s_1(t)+(k-s_1(t))^+)\rmd t$.    
\label{lem:waterfilling}
\end{lemma}

\begin{proof}
It is sufficient to prove \cref{clm:kkt}, which will be proved by the KKT condition.
\begin{claim}
Given a function $s_1:[0,1]\to [0,+\infty)$ and a constant $M\geq 0$, consider the following maximization problem:
\[
\max_{s_2:[0,1]\to[0,\infty)} J(s_2):=\int_{0}^{1}\ln(s_1(t)+s_2(t))\rmd t \quad\text{ s.t. }\quad \int_{0}^{1}s_2(t)\rmd t = M \text{ and } s_2(t)\geq 0.
\]
Then, $J(s_2)$ is maximized when $s_2(t)=\max\{0,k-s_1(t)\}$ for some fixed $k\geq 0$.
\label{clm:kkt}
\end{claim}
\begin{proofof}{\cref{clm:kkt}}
We rewrite the problem as minimizing $-J(s_2)$ subject to $\int_{0}^{1} s_2(t)\rmd t=M$ and $-s_2\le 0$.
The Lagrangian is
\[
\cL(s_2,\lambda,\mu)
=
-\int_0^1 \ln\big(s_1+s_2\big)\rmd t
+\lambda\Big(\int_0^1 s_2\rmd t - M\Big)
+\int_0^1 \mu(t)\,(-s_2(t))\rmd t,
\]
where $\mu\ge 0$.  
Since $-J$ is convex, the constraints are convex, the KKT conditions are necessary and sufficient, and yield global optimality.
Taking the derivation $\cL$ for $s_2$ gives,
\[
-\frac{1}{s_1(t)+s_2^\star(t)}+\lambda-\mu(t)=0
\quad\Longleftrightarrow\quad
\frac{1}{s_1(t)+s_2^\star(t)}=\lambda-\mu(t),
\]
where $s^*_2$ is the function that maximizes $J(s_2)$.
The complementary slackness enforces $\mu(t)\,s_2^*(t)=0$. Hence:
\begin{itemize}
\item If $s_2^*(t)>0$, then $\mu(t)=0$ and $s_1(t)+s_2^*(t)=1/\lambda$.
\item If $s_2^*(t)=0$, then $\mu(t)=\lambda-\frac{1}{s_1(t)}\ge 0$, so $s_1(t)\ge 1/\lambda$.
\end{itemize}
Let $k:=1/\lambda>0$. The structure above is exactly
\[
s_2^*(t)=\max\{k-s_1(t),0\}=(k-s_1(t))^+ .
\]
This proves \cref{clm:kkt}.
\end{proofof}

Applying \cref{clm:kkt} directly proves \cref{lem:waterfilling}.
\end{proof}

\subsection{Final Analysis Using a Computer Program}
\label{subsec:computer-program}

We first simplify \mpref{mp:one} to obtain \mpref{mp:two} based on the worst-case transformation based on \cref{subsec:unique-t}-\ref{subsec:waterfilling}.
And then, we analyze \mpref{mp:two} using a computer program.

\subsubsection{Obtaining \mpref{mp:two} based on \cref{subsec:unique-t}-\ref{subsec:waterfilling}}
\label{subsec:mp2}

We first use the water-filling bound from \cref{subsec:waterfilling} to simplify the value of the input distribution in \eqref{equ:ratio-1}.
Recall that $\bar\mu^{\sm}=(1-1/e)\E[S^{\sm}]$ and $\bar\mu^{\md}=(1-1/e)\E[S^{\md}]$.
Then, \cref{lem:waterfilling} implies 
\[
\E\left[\val(S^{\md})+\left(\frac{e}{e-1}\right)\cdot \val(S^{\sm})\right]=\E[(k-\val(S^{\lg}))^+]=\left(\frac{e}{e-1}\right)\cdot \bar\mu^{\md}+\left(\frac{e}{e-1}\right)^2\cdot\bar\mu^{\sm}.
\]
Also note that $T^{\lg}$ and $S^{\lg}$ are identical distributions by \cref{subsec:s-t-large-same}.
Thus, we have
\begin{align*}
&\E\left[\ln\left(\val(S^{\lg})+\val(S^{\md})+\frac{e}{e-1}\cdot \val(S^{\sm})\right)\right]=\E\left[\ln\left(\val(S^{\lg})+(k-\val(S^{\lg}))^+\right)\right] \\
&\leq \E\left[\ln\left(\max\{\val(S^{\lg}),k\}\right)\right]=\E\left[\max\{\ln (\val(S^{\lg})),\ln k\}\right]=\E\left[\max\{\ln (\val(T^{\lg})),\ln k\}\right].
\end{align*}
From now on, we shall drop the notation $S^{\lg}$.
From \cref{subsec:align:1} and \cref{subsec:align:2}, we have $\val(T^{\lg})=t^{\nlg}\geq 1$ with probability $1$. 
Hence, we have 
\begin{align*}
\eqref{equ:ratio-1}&\leq\E\left[\max\{\ln (\val(T^{\lg})),\ln k\}\right]-\E\left[\ln\left(\max\left\{\val(T^{\lg}) , t^{\nlg} \right\} \right)\right] \\
&=
\E\left[\max\{\ln (t^{\nlg}),\ln k\}\right]-\E\left[\ln\left(t^{\nlg}\right)\right] 
=\E\left[\left(\ln k -\ln t^{\nlg}\right)^+\right].
\end{align*}
Again due to $\val(T^{\lg})=t^{\nlg}\geq 1$ with probability $1$ and \cref{subsec:s-t-large-same}, \cref{lem:waterfilling} implies:
\[
\E\left[\val(S^{\md})+\frac{e}{e-1}\cdot\val(S^{\sm})\right]=\int_{0}^{1}(k-s_1(t))^+\rmd t=\E\left[\left(k-t^{\nlg}\right)^+\right].
\]

In summary, \mpref{mp:one} is upper bounded by \mpref{mp:two}.

\mppara{2}{mp:two}{
Given a fixed sufficiently small constant ${\epsilon_1}>0$. We aim to maximize:
\begin{align}
    \E\left[\left(\ln k -\ln t^{\nlg}\right)^+\right],
    \label{equ:ratio-2}
\end{align}
subject to the following constraints:
\begin{enumerate}[label=(MP2\alph*),leftmargin=*]
    \item $t^{\nlg}\in\R_{\geq 0}$ is a random value and $\bar\mu^{\md},\bar\mu^{\sm},k\geq 0$ are real numbers.
    \item $\Pr[t^{\nlg}\geq 1]=1$.
    \item $\E[t^{\nlg}]\ge \bar\mu^{\md}+\bar\mu^{\sm}$.
    \label{mp2:key}
    \item For any $\lambda<0$, $\E\left[e^{\lambda\cdot t^{\nlg}}\right]\le e^{(e^{\lambda}-1)\bar\mu^{\md}+(e^{\lambda\sqrt{\epsilon_1}}-1)\bar\mu^{\sm}/\sqrt{{\epsilon_1}}}$.
    \label{mp2:concentration}
    \item $\E[(k-t^{\nlg})^+]= \left(\frac{e}{e-1}\right)\cdot \bar\mu^{\md}+\left(\frac{e}{e-1}\right)^2\cdot\bar\mu^{\sm}$.
    \label{mp2:k}
\end{enumerate}}

\subsubsection{Analysis \mpref{mp:two} with a Computer Program}
\label{subsec:code}
For simplicity, we let $\bar\mu:=\bar\mu^{\md}+\bar\mu^{\sm}$, and define $\alpha:=\bar\mu^{\sm}/\bar\mu$. 
Thus, our program becomes:

\mppara{3}{mp:three}{
Given a fixed sufficiently small constant ${\epsilon_1}>0$, we aim to maximize
\begin{align}
\E\left[\left(\ln k -\ln t^{\nlg}\right)^+\right],
\label{equ:ratio-4}
\end{align}
subject to the following constraints:
\begin{enumerate}[label=(MP3\alph*),leftmargin=*]
    \item $t^{\nlg}$ is a random value and $\bar\mu\ge 0,k\geq 0, 0\le \alpha\le 1$ are real numbers.
    \item $\Pr[t^{\nlg}\ge 1]=1$.
    \label{mp3:t}
    \item $\E[t^{\nlg}]\ge \bar\mu$.
    \label{mp3:mu}
    \item For any $\lambda<0$, $\E\left[e^{\lambda\cdot t^{\nlg}}\right]\le e^{(e^{\lambda}-1)(1-\alpha)\bar\mu+(e^{\lambda\sqrt{\epsilon_1}}-1)(\alpha\bar\mu)/\sqrt{{\epsilon_1}}}$.
    \label{mp3:concentration}
    \item $\E[(k-t^{\nlg})^+]= \left(\frac{e}{e-1}\right)\cdot(1-\alpha)\bar\mu+\left(\frac{e}{e-1}\right)^2\cdot\alpha\bar\mu$.
    \label{mp3:k}
\end{enumerate}}

Our goal then is to show that the value of \mpref{mp:three} is at most $\ln(3.56)$.

\paragraph{Reducing the search space.} Before running the computer program, we first show in \cref{lem:large-mu} that when $\bar\mu$ or $k$ is very large, the value of \mpref{mp:three} is small, and then, show in \cref{lem:small-mu} that when $\bar\mu\leq 1$, the value of \mpref{mp:three} is also small.
This reduces the search space of the computer program.
When $\bar\mu$ is large, we directly use the Chernoff bound to bound the objective in this case.

Before showing these two lemmas, we first relax \ref{mp3:concentration} and \ref{mp3:k} by setting $\alpha=0$ and $\alpha=1$ in these two inequalities, respectively.
This provides us with an upper bound of $\E[\exp(\lambda t^{\nlg})]$ and $\E[(k-t^{\nlg})^+]$, and thus increases the value of \eqref{equ:ratio-4}:
\begin{equation}
\E\left[e^{\lambda\cdot t^{\nlg}}\right]\le e^{(e^{\lambda}-1)\bar\mu},\forall \lambda<0 \quad\text{and}\quad \E[(k-t^{\nlg})^+]\leq \left(\frac{e}{e-1}\right)^2\cdot\bar\mu.
\label{equ:mp4-concentration-k}
\end{equation}
For simplicity, in the proof of \cref{lem:small-mu} and \cref{lem:large-mu}, we use $\mu$ and $t$ to represent $\bar\mu$ and $t^{\nlg}$.
Observe that
\begin{equation}
(\ln k - \ln t)^+=\ln\left(\frac{(k-t)^+}{t}+1\right)
\label{equ:k-t}
\end{equation}

We are now ready to prove \cref{lem:small-mu} and \cref{lem:large-mu}.

\begin{lemma}
In \mpref{mp:three}, when $\bar\mu\leq 1$, the value of \eqref{equ:ratio-4} is at most $\ln(1+(e/(e-1))^2)$.
\label{lem:small-mu}
\end{lemma}

\begin{proof}
By \eqref{equ:k-t}, we have the following inequalities for \eqref{equ:ratio-4}:
\begin{align*}
\E\left[\ln\left(\frac{(k-1)^+}{t}+1\right)\right]\leq \E\left[\ln\left((k-t)^++1\right)\right] \leq \ln\left(\E[(k-t)^+]+1\right)\leq \ln\left(\left(\frac{e}{e-1}\right)^2+1\right),
\end{align*}
where the first inequality is due to $t\geq 1$ \ref{mp3:t}, the second inequality is due to Jensen's inequality, and the last inequality is due to \eqref{equ:mp4-concentration-k} and $\bar\mu \leq 1$.
\end{proof}

\begin{lemma}
    In \mpref{mp:three}, when $\bar\mu\geq 3.6\times 10^5$, the value of \eqref{equ:ratio-4} is at most $\ln(1+(e/(e-1))^2)+0.015982$.
\label{lem:large-mu}
\end{lemma}

\begin{proof}

We apply the standard proof of Chernoff bound~\cite[Lemma C.1]{arxiv/abs-0909-4348}, the concentration bound in \eqref{equ:mp4-concentration-k} together with \ref{mp3:mu} implies the following low-tail bound:
\begin{equation}
\Pr\left[t\leq (1-\delta)\mu\right]\leq e^{-\delta^2\mu/2},\quad \text{for all }\delta\in(0,1).
\label{equ:ratio-4:large-chernoff}
\end{equation}
By \eqref{equ:k-t}, \eqref{equ:ratio-4} is equal to the summation of the following two terms:
\begin{fleqn}[0pt]
    \begin{align}
    &\Pr\left[t\geq (1-\delta)\mu\right]\cdot\E\left[ \ln\left(\frac{(k-t)^+}{t}+1\right)\Bigg|t\geq (1-\delta)\mu\right]; \label{equ:ratio-4:right} \\
    &\Pr\left[t\leq (1-\delta)\mu\right]\cdot\E\left[ \ln\left(\frac{(k-t)^+}{t}+1\right)\Bigg|t\leq (1-\delta)\mu\right]. \label{equ:ratio-4:left}
    \end{align}
\end{fleqn}
In the following, we bound \eqref{equ:ratio-4:right} and \eqref{equ:ratio-4:left}, respectively.
For \eqref{equ:ratio-4:left}, we apply the Chernoff bound \eqref{equ:ratio-4:large-chernoff} directly:
\begin{align*}
&\Pr\left[t\leq (1-\delta)\mu\right]\cdot\E\left[ \ln\left(\frac{(k-t)^+}{t}+1\right)\Bigg|t\leq (1-\delta)\mu\right] \\
&\leq \frac{\E\left[ \ln\left((k-t)^++1\right)|t\leq (1-\delta)\mu\right]}{e^{\delta^2\mu/2}} \tag{\eqref{equ:ratio-4:large-chernoff} and $t\geq 1$} \\
&\leq \frac{\ln\left(\E[(k-t)^+|t\leq (1-\delta^2)\mu]+1\right)}{e^{\delta^2\mu/2}} \tag{Jensen's inequality} \\
&\leq \frac{\ln\left(1+\left(\frac{e}{e-1}\right)^2\mu\right)}{e^{\delta^2\mu/2}}:=y(\mu)\tag{\eqref{equ:mp4-concentration-k}}
\end{align*}
Now, we upper bound the value of \eqref{equ:ratio-4:right}.
\begin{align*}
&\Pr\left[t\geq (1-\delta)\mu\right]\cdot\E\left[ \ln\left(\frac{(k-t)^+}{t}+1\right)\Bigg|t\geq (1-\delta)\mu\right] \\
&\leq \left(1-e^{-\delta^2\mu/2}\right)\cdot\E\left[ \ln\left(\frac{(k-t)^+}{(1-\delta)\mu}+1\right)\Bigg|t\geq (1-\delta)\mu\right] \tag{\eqref{equ:ratio-4:large-chernoff} and $t\geq (1-\delta)\mu$} \\
&\leq \left(1-e^{-\delta^2\mu/2}\right)\cdot\ln\left(1+\frac{\left(\frac{e}{e-1}\right)^2}{1-\delta}\right) \tag{Jensen's inequality and \eqref{equ:mp4-concentration-k}}\\
&\leq \ln\left(1+\frac{\left(\frac{e}{e-1}\right)^2}{1-\delta}\right) \tag{$\left(1-e^{-\delta^2\mu/2}\right)\leq 1$}
\end{align*}
Note that for any $0<x\leq 1/2$ and a constant $A$, we have
\[
\ln\left(1+\frac{A}{1-x}\right)=\ln(1+A-x)-\ln(1-x)\leq \ln(1+A)+\frac{x}{1-x} \leq \ln(1+A)+2x,
\]
where the first inequality is due to $x>0$ and $-\ln(1-x)\leq x/(1-x)$ for any $x\in(0,1)$, the second inequality is due to $x\leq 1/2$.
Later, we will choose a $\delta<1/2$, and thus, we have
\[
\eqref{equ:ratio-4:right} \leq \ln\left(1+\frac{\left(\frac{e}{e-1}\right)^2}{1-\delta}\right) \leq \ln\left(1+\left(\frac{e}{e-1}\right)^2\right)+2\delta.
\]
In summary, we have
\begin{equation}
\eqref{equ:ratio-4}=\E\left[\left(\ln k -\ln t^{\nlg}\right)^+\right]\leq \ln\left(1+\left(\frac{e}{e-1}\right)^2\right)+2\delta+y(\mu).
\label{equ:ratio-4:large}
\end{equation}
We need to set up appropriate $\mu$ and $\delta$ such that both $\delta$ and $y(\mu)$ are super small.
One can easily verify that the following inequality holds if $\mu\geq3.6\times 10^5$ and $\delta=0.0079$ (note that $y(\mu)$ decreases in $[3.6\times 10^5,+\infty)$):
\[
y(\mu)<0.000182
\quad\text{and}\quad
\exp\left(\ln\left(1+\left(\frac{e}{e-1}\right)^2\right)+2\delta+y(\mu)\right) < 3.55908.
\]
This proves the lemma since $2\delta+y(\mu)<0.015982$.
\end{proof}

When $\bar\mu$ is small, we use the following lemma to bound the range of $k$.

\begin{lemma}
    \label{lem:bound-k}
    When $\bar\mu$ and $k$ are fixed, the value of \mpref{mp:three} is at most the following term:
    \begin{align*}
        \left(\frac{e}{e-1}\right)^2\cdot\frac{\bar\mu\ln k}{k}.
    \end{align*}
\end{lemma}

\begin{proof}
    We bound the objective as
    \begin{align*}
        &\E\left[[1\le t^{\nlg}< k]\cdot \left(\ln k -\ln t^{\nlg}\right)\right] \\
        ={}&\E\left[[1\le t^{\nlg}< k]\cdot \left(\ln k -\ln t^{\nlg}\right)\big/\left( k -t^{\nlg}\right)\cdot \left( k -t^{\nlg}\right)\right] \\
        \le{}&\max_{1\le  t^{\nlg}< k}\left\{\left(\ln k -\ln t^{\nlg}\right)\big/\left( k -t^{\nlg}\right)\right\}\cdot \E\left[\left( k -t^{\nlg}\right)^+\right].
    \end{align*}
    Since $\left(\ln k -\ln t^{\nlg}\right)\big/\left( k -t^{\nlg}\right)$ is the slope of the line passing through both $(k,\ln k)$ and $(t,\ln t)$, and $\ln x$ is a concave function, the maximum of the first term is attained when $t^{\nlg}=1$. For the second term, we directly use \ref{mp3:k} to bound it as $(e/(e-1))^2\cdot\bar\mu$ by taking $\alpha=1$. Thus, the objective is at most
    \begin{align*}
        \frac{(\ln k)}{k}\cdot \left(\frac{e}{e-1}\right)^2\cdot\bar\mu.
    \end{align*}
\end{proof}

Next, we relax \mpref{mp:three} into a program that is easy to solve using a computer.

\paragraph{Modifying the program: Fixing the parameters.} We first fix a range for each of the three parameters $k,\bar\mu,\alpha$, denoted as $[k_L,k_R],[\bar\mu_L,\bar\mu_R]$ and $[\alpha_L,\alpha_R]$. The value of the following program upper bounds the value of \mpref{mp:three} when we add the constraints $k\in [k_L,k_R]$, $\bar\mu\in [\bar\mu_L,\bar\mu_R]$ and $\alpha\in [\alpha_L,\alpha_R]$:

\mppara{3-1}{mp:three-alt}{
Given a fixed sufficient small constant ${\epsilon_1}>0$, we aim to maximize
\begin{align}
\E\left[\left(\ln k_R -\ln t^{\nlg}\right)^+\right],
\label{equ:ratio-4-1}
\end{align}
subject to the following constraints:
\begin{enumerate}[label=(MP3-1\alph*),leftmargin=*]
    \item $t^{\nlg}\geq 0$ is a random variable.
    \item $\Pr[t^{\nlg}\ge 1]=1$.
    \item $\E[t^{\nlg}]\ge \mu_L$.
    \item For any $\lambda<0$, $\E\left[e^{\lambda\cdot t^{\nlg}}\right]\le e^{(e^{\lambda}-1)(1-\alpha_L)\bar\mu_L+(e^{\lambda\sqrt{\epsilon_1}}-1)(\alpha_L\bar\mu_L)/\sqrt{{\epsilon_1}}}$.
    \item $\E[(k_L-t^{\nlg})^+]\le  \left(\frac{e}{e-1}\right)\cdot(1-\alpha_R)\bar\mu_R+\left(\frac{e}{e-1}\right)^2\cdot\alpha_R\bar\mu_R$.
\end{enumerate}}

The benefit of considering \mpref{mp:three-alt} is that it is a linear program in the variables $y_t=\Pr[t^{\nlg}=t]$ $(t\in \R)$. 

\paragraph{Modifying the program: Discretizations.} However, we still have an infinite number of variables and constraints, which is unacceptable. Next, we reduce both of these numbers.
\begin{itemize}
    \item \textbf{Discretize the variable $t^{\nlg}$:} Let $l_t$ be an integer parameter to be determined later, let $t_1=1,\dots,t_{l_t+1}=k_R$ be evenly spaced values, and let $x_i$ $(1\le i\le l_t)$ be a variable indicating the probability that $t^{\nlg}$ is in $[t_i,t_{i+1})$ (the cases where $t^{\nlg}>k_R$ does not matter to us much). We discretize the program by only considering the variables $x_i$.
    \item \textbf{Discretize the constraints:} Let $l_\lambda$ be an integer parameter to be determined later, and let $\lambda_j=\ln(1-j/(l_\lambda+1))$ for $j=1\dots l_\lambda$. Instead of considering every $\lambda<0$, we only consider the $\lambda_j$'s. In the actual computer program, we always have $l_\lambda=60$. We can let ${\epsilon_1}$ be a sufficiently small number such that $e^{\lambda_j\sqrt{{\epsilon_1}}-1}/\sqrt{{\epsilon_1}}\ge \lambda_j+0.0001$ for any $1\le j\le l_\lambda$.
\end{itemize}
After discretization, we obtain the following \emph{linear} program whose value upper bounds that of \mpref{mp:three-alt}:

\mppara{3-2}{mp:three-alt-1}{Maximize
\begin{align}
\sum_{i=1}^{l_t}\left(\ln k_R -\ln t_i\right)^+\cdot x_i,
\label{equ:ratio-4-2}
\end{align}
subject to the following constraints:
\begin{enumerate}[label=(MP3-2\alph*),leftmargin=*]
    \item $x_i\ge 0$ for $1\le i\le l_t$.
    \item $\sum_{i=1}^{l_t}x_i\le 1$.
    \item For any $1\le j\le l_\lambda$, $\sum_{i=1}^{l_t}e^{\lambda_j\cdot t_{i+1}}\cdot x_i\le e^{(e^{\lambda_j}-1)(1-\alpha_L)\bar\mu_L+(\lambda_j+0.0001)(\alpha_L\bar\mu_L)}$.
    \item $\sum_{i=1}^{l_t}(k_L-t_{i+1})^+\cdot x_i\le  \left(\frac{e}{e-1}\right)\cdot(1-\alpha_R)\bar\mu_R+\left(\frac{e}{e-1}\right)^2\cdot\alpha_R\bar\mu_R$.
\end{enumerate}}

We further relax (MP4d) to
\begin{align*}
    \sum_{i=1}^{l_t}\min\{10000,e^{\lambda_j\cdot t_{i+1}-(e^{\lambda_j}-1)(1-\alpha_L)\bar\mu_L-(\lambda_j+0.0001)(\alpha_L\bar\mu_L)}\cdot x_i\}\le 1
\end{align*}
to avoid precision errors when running the computer program.

\paragraph{Running the computer program.} The computer program uses a divide-and-conquer approach: It starts with the intervals $[\bar\mu_L,\bar\mu_R]$, $[k_L,k_R]$ and $[\alpha_L,\alpha_R]$, and checks whether the value of \mpref{mp:three-alt-1} under these parameters is at most $\ln(3.56)>1.269$. If not, the program subdivides one of the intervals and recurses.

We run the computer program on the following intervals:
\begin{itemize}
    \item When $1\le \bar\mu\le 3$, using \cref{lem:bound-k}, we only need to check the case where $k\le 20$.
    \item When $3<\bar\mu\le 3.6\times 10^5$, using \cref{lem:bound-k}, we only need to check the case where $k\le 2\times 10^7$.
\end{itemize}

Our computer program is written in MATLAB and can be found at the following link:
\begin{itemize}
    \item \href{https://github.com/ruilong-zhang/WeightNashSocialWelfare}{https://github.com/ruilong-zhang/WeightNashSocialWelfare}
\end{itemize}

\paragraph{Remark.} The optimal value of \mpref{mp:three} is strictly larger than $\ln(1+(e/(e-1))^2)$. Consider the following solution to \mpref{mp:three}: $\Pr[t^\nlg=1.0673]=p:=0.936, \Pr[t^\nlg=50]=1-p, (1-\alpha)\bar\mu=0.21, \alpha\bar\mu=1$, then $k\approx4.096$ and one can verify that this is indeed a feasible solution to \mpref{mp:three} with an objective value strictly larger than $1.2588>\ln(1+(e/(e-1))^2)$.
Thus, the ratio is strictly larger than $e^{1.2588}>3.521>1+(e/(e-1))^2$.

\subsection{Handling Large $p^*$ Case}
\label{subsec:large-p*}

Recall that in this case, all items in $M^{\nlg}$ are well concentrated since $\E[\val(S^{\md})+\val(S^{\sm})]=\Omega(1/{\epsilon_1}^{1/6})$ by \cref{lem:upper-bound-input-large}.
Therefore, it is unnecessary to further differentiate between medium and small items. 
A mathematics program is still required to bridge the gap between $\vlp$ and $\valg$. 
Although the worst-case reductions discussed in the previous section are still needed, the help of a computer program is no longer necessary.

Since we no longer distinguish medium and small items, the Chernoff bound proved by~\cite{stoc/FHLZ25} is sufficient for us.
Recall that $\bx$ is the LP solution and $\bx^{\nlg}$ is the vector $\bx$ that is restricted to the items in $M^{\nlg}$.
Recall that $T$ is the item set that is assigned to the agent by the rounding algorithm and $T^{\nlg}=T\cap M^{\nlg}$.

\begin{lemma}[\cite{stoc/FHLZ25}]
Let $\mu^{\nlg}:=F(\bx^{\nlg})$ and for any $\delta\in(0,1)$, we have
\[
\Pr[\val(T^{\nlg})\leq (1-\delta)\cdot\mu^{\nlg}] \leq e^{-\delta^2\mu^{\nlg}/2}.
\]
\label{lem:chernoff}
\end{lemma}

Similar to \cref{lem:upper-bound-input-small}, we first develop an upper bound of $\vlp$ (\cref{lem:large-p:vlp-upper}).
\begin{lemma}
Fix a sufficiently small ${\epsilon_1}>0$, we have the following upper bound of $\vlp$:
\[
\vlp\leq \E\left[\ln\left(\val(S^{\lg})+\frac{e}{e-1}\cdot \val(S^{\nlg})\right)\right]+O({\epsilon_1})
\]
\label{lem:large-p:vlp-upper}
\end{lemma}
\begin{proof}
Similar to \cref{lem:upper-bound-input-small}, we compare $\ln(\val(S^{\enu})+e/(e-1)\cdot \val(S^{\nenu}))$ and $\ln(\val(S^{\lg})+e/(e-1)\cdot \val(S^{\nlg}))$.
The former is at most the latter if $S^{\nenu}\subseteq S^{\nlg}$.
Otherwise, we have $S^{\lg}\cap S^{\nenu}\ne\emptyset$, and the former is at most the latter plus $\ln(e/(e-1))=\Theta(1)$.
The lemma follows from the fact that there can only be ${\epsilon_1}$ fraction of configurations $S$ with $S^{\lg}\cap S^{\nenu}\ne\emptyset$ (because we have only one unit of large items).
\end{proof}

Similarly to the previous section, applying the lower bound of the output distribution \eqref{equ:lower-bound-output}, we then aim to bound the difference of \eqref{equ:large-p:ratio-1}, where we again ignore the factor ${\epsilon_1}$ from \cref{lem:large-p:vlp-upper} for the moment.
\begin{equation}
\E\left[\ln\left(\val(S^{\lg})+\frac{e}{e-1}\cdot \val(S^{\nlg})\right)\right]-\E\left[\ln\left(\max\left\{  \val(T^{\lg}) , t^{\nlg} \right\} \right)\right].
\label{equ:large-p:ratio-1}
\end{equation}

Applying the worst-case reduction stated in the previous sections, we can assume all properties in \cref{cor:large-p:assumption}:
\begin{corollary}
Assuming the following properties does not decrease the value of \eqref{equ:large-p:ratio-1}:
\begin{enumerate}[leftmargin=*,label=(\ref{cor:large-p:assumption}\ablue{\alph*})]
    \item $S^{\lg}$ and $T^{\lg}$ are the same distributions.
    \label{prop:large-p:same}
    \item We have $\val(T^{\lg})= t^{\nlg} \geq 1$.
    \label{prop:large-p:align}
    \item There exists a $k\geq 0$ such that the following inequalities hold:
    \begin{fleqn}[0pt]
    \begin{align*}
    &\E\left[\left(\frac{e}{e-1}\right)\cdot \val(S^{\nlg})\right]=\E[(k-\val(S^{\lg})^+]; \\
    &\E\left[\ln\left(\val(S^{\lg})+\left(\frac{e}{e-1}\right)\cdot \val(S^{\nlg})\right)\right]\leq \E\left[\ln\left(\val(S^{\lg})+(k-\val(S^{\lg}))^+\right)\right]. 
    \end{align*}
    \end{fleqn}
    \label{prop:large-p:waterfilling}
\end{enumerate}
\label{cor:large-p:assumption}
\end{corollary}
Note that by \ref{property:rounding-lp} and \ref{prop:large-p:waterfilling}, we have
\begin{fleqn}[0pt]
    \begin{align}
    &\E\left[\left(\frac{e}{e-1}\right)\cdot \val(S^{\nlg})\right] \leq \left(\frac{e}{e-1}\right)^2\cdot\mu^{\nlg}; \label{equ:large-p:k-vs} \\
    &\E\left[\ln\left(\val(S^{\lg})+(k-\val(S^{\lg}))^+\right)\right]\leq \E\left[\max\left\{\ln(\val(S^{\lg})),\ln(k)\right\}\right]. \label{equ:large-p:objective}
    \end{align}
\end{fleqn}
Thus, by \ref{prop:large-p:same}, \ref{prop:large-p:align}, \ref{prop:large-p:waterfilling} and \eqref{equ:large-p:objective}, we again have:
\[
\eqref{equ:large-p:ratio-1}\leq \E\left[\max\{\ln (\val(T^{\lg})),\ln k\}\right]-\E\left[\ln\left(\max\left\{  \val(T^{\lg}) , t^{\nlg} \right\} \right)\right]=\E\left[\left(\ln k -\ln t^{\nlg}\right)^+\right].
\]

In summary, we get \mpref{mp:four} for this case.
\mppara{4}{mp:four}{
Given a fixed sufficiently small constant ${\epsilon_1}>0$, our goal is to maximize
\begin{equation}
\E\left[\left(\ln k-\ln t^{\nlg}\right)^+\right],
\label{equ:ratio-3}
\end{equation}
subject to the following constraints:
\begin{enumerate}[label=(MP4\alph*),leftmargin=*]
    \item $t^{\nlg}\geq 1$ is a random variable and $k\geq 0$ is a real number.
    \label{mp4:1}
    \item $\mu^{\nlg}=\Omega(1/{\epsilon_1}^{1/6})$.
    \label{mp4:mu}
    \item For any $\delta\in(0,1)$, $\Pr[t^{\nlg}\leq (1-\delta)\cdot\mu^{\nlg}] \leq e^{-\delta^2\mu^{\nlg}/2}$.
    \label{mp4:concentration}
    \item $\E[(k-t^{\nlg})^+] \leq \left(\frac{e}{e-1}\right)^2\cdot \mu^{\nlg}$.
    \label{mp4:k}
\end{enumerate}
}
In \mpref{mp:four}, \ref{mp4:mu} is due to \cref{lem:upper-bound-input-large}, \ref{mp4:concentration} is due to \cref{lem:chernoff}, and \ref{mp4:k} is due to \eqref{equ:large-p:k-vs} and \ref{prop:large-p:waterfilling}.
The rest of \cref{subsec:large-p*} aims to find an upper bound of \mpref{mp:four}.
For simplicity, we shall drop the script ``$\nlg$'' from the notations, i.e., we use $t$ and $\mu$ to represent $t^{\nlg}$ and $\mu^{\nlg}$ in the remainder of \cref{subsec:large-p*}.
Then, we have \cref{lem:large-p:ratio} whose proof is essentially the same as \cref{lem:large-mu}.

\begin{lemma}
The value of \eqref{equ:ratio-3} in \mpref{mp:four} is at most $\ln(1+(e/(e-1))^2)+O(\epsilon)$. 
\label{lem:large-p:ratio}
\end{lemma}

\begin{proof}
Similar to \cref{lem:large-mu}, we have $(\ln k - \ln t)^+=\ln\left((k-t)^+/t+1\right)$.
Then, the remaining proof is the same as \cref{lem:large-mu} where we only need to set up different parameters.
In the following, we only sketch the key inequalities.
Note that \eqref{equ:ratio-3} is again equal to the summation of the following two terms:
\begin{fleqn}[0pt]
    \begin{align}
    &\Pr\left[t\geq (1-\delta)\mu\right]\cdot\E\left[ \ln\left(\frac{(k-t)^+}{t}+1\right)\Bigg|t\geq (1-\delta)\mu\right]\leq \frac{\ln\left(1+\left(\frac{e}{e-1}\right)^2\mu\right)}{e^{\delta^2\mu/2}}:=y(\mu); \label{equ:ratio-3:right} \\
    &\Pr\left[t\leq (1-\delta)\mu\right]\cdot\E\left[ \ln\left(\frac{(k-t)^+}{t}+1\right)\Bigg|t\leq (1-\delta)\mu\right]\leq \ln\left(1+\left(\frac{e}{e-1}\right)^2\right)+2\delta, \label{equ:ratio-3:left}
    \end{align}
\end{fleqn}
where \eqref{equ:ratio-3:right} applies \ref{mp4:concentration}, \eqref{equ:large-p:k-vs} and \ref{mp4:1}, while \eqref{equ:ratio-3:left} applies \ref{mp4:concentration} and \eqref{equ:large-p:k-vs}.
We again need to choose a $\delta<1/2$ to get \eqref{equ:ratio-3:left}.
In summary, we again have:
\[
\eqref{equ:ratio-3}=\E\left[\left(\ln k-\ln t^{\nlg}\right)^+\right] \leq \ln\left(1+\left(\frac{e}{e-1}\right)^2\right)+y(\mu)+2\delta.
\]

We set $\delta = \epsilon_1^{1/20} = \epsilon$. As we assumed $\mu \geq c \cdot (1/\epsilon_1)^{1/6} = c\cdot \epsilon^{-10/3}$, we have 
\begin{align*}
    \frac{\ln\left(1+\left(\frac{e}{e-1}\right)^2\mu\right)}{e^{\delta^2\mu/2}} \leq \frac{O(\mu)}{e^{\epsilon^2\mu/2}} \leq O(\epsilon).
\end{align*}

On the other hand, we have $2\delta=2{\epsilon_1}^{1/20} = 2\epsilon$.
Hence, \eqref{equ:ratio-3} is at most $\ln(1+(e/(e-1))^2)+O(\epsilon)$, which proves \cref{lem:large-p:ratio}.
\end{proof}

\subsection{Proof of \cref{lem:ratio}}
\label{subsec:proofofratio}

With the \cref{subsec:computer-program} and \cref{subsec:large-p*}, we are ready to prove \cref{lem:ratio}.

\begin{proofof}{\cref{lem:ratio}}
For the small $p^*$ case, by \cref{lem:upper-bound-input-small}, we have
\[
\vlp \leq \E\left[ \ln\left(\val(S^{\lg})+\val(S^{\md})+\frac{e}{e-1}\cdot \val(S^{\sm}) \right)\right] +O({\epsilon_1}^{1/3}).
\]
From \cref{subsec:code}, we know that \mpref{mp:three} is at most $\ln(3.56)$. 
Since \mpref{mp:three} upper bounds \mpref{mp:one}, we have:
\[
\E\left[\ln\left(\val(S^{\lg})+\val(S^{\md})+\frac{e}{e-1}\cdot \val(S^{\sm})\right)\right]-\E\left[\ln\left(\max\left\{  \val(T^{\lg}) , t^{\nlg} \right\} \right)\right] \leq \ln(3.56).
\]
Note that $\vlp\geq \E\left[\ln\left(\max\left\{  \val(T^{\lg}) , t^{\nlg} \right\} \right)\right]$ by \eqref{equ:lower-bound-output}.
Hence, we have
\[
\vlp-\valg \leq \ln(3.56)+O({\epsilon_1}^{1/3})=\ln(3.56)+O({{\epsilon}}),
\]
since ${{\epsilon}}={\epsilon_1}^{1/20}\geq{\epsilon_1}^{1/3}$ for sufficiently small ${\epsilon_1}$.

For the large $p^*$ case, since \mpref{mp:four} upper bounds \eqref{equ:large-p:ratio-1}, by \cref{lem:large-p:ratio}, we know that 
\[
\E\left[\ln\left(\val(S^{\lg})+\frac{e}{e-1}\cdot \val(S^{\nlg})\right)\right]-\E\left[\ln\left(\max\left\{  \val(T^{\lg}) , t^{\nlg} \right\} \right)\right] \leq \ln\left(1+\left(\frac{e}{e-1}\right)^2\right)+O({\epsilon_1}^{1/20}).
\]
By \cref{lem:large-p:vlp-upper}, we have
\[
\vlp\leq \E\left[\ln\left(\val(S^{\lg})+\frac{e}{e-1}\cdot \val(S^{\nlg})\right)\right]+O({\epsilon_1}),
\]
Again, by \eqref{equ:lower-bound-output}: $\vlp\geq \E\left[\ln\left(\max\left\{  \val(T^{\lg}) , t^{\nlg} \right\} \right)\right]$, we have
\[
\vlp-\valg \leq \ln\left(1+\left(\frac{e}{e-1}\right)^2\right)+O({\epsilon_1}^{1/20})+O({\epsilon_1})\leq \ln(3.56)+O({{\epsilon}}).
\]
Combining these two cases proves \cref{lem:ratio}.
\end{proofof}

\subsection{The Upper Bound of Integrality Gap}
\label{subsec:upper-integap}

This section aims to show the upper bound of the integrality gap of \eqref{Conf-LP}.
Note that the approximation ratio in \cref{thm:ratio} is not the actual upper bound because we also lose some factor on solving LP (i.e. \cref{lem:lp-solver}).
To obtain the upper bound of the integrality gap, we assume that we are given an optimal fractional solution $\by^*:=(y^*_{i,S})_{i\in N, S\subseteq M}$ to \eqref{Conf-LP}.

We then construct a bipartite graph $G:=(N\cup M, W)$ with an edge value vector $\bx\in[0,1]^{\abs{E}}$ and then apply the rounding algorithm stated in \cref{subsec:rounding} to obtain a randomized allocation.
We can also assume that the configurations of $\by^*$ form a partition of $M$, and that an item is either large or small in the rounding algorithm (i.e. $G$ is a simple graph and there are no parallel edges).
Similarly, fix an arbitrary agent $i\in N$, we can define the input and output distribution as follows:
\[
\vlp=\E[\ln(\val(S))]=\E[\ln(\val(S^{\lg})+\val(S^{\nlg}))] \qquad\text{and}\qquad \valg\geq\E\left[\ln\left(\max\left\{\val(T^{\lg}),T^{\nlg}\right\}\right)\right],
\]
where the input distribution becomes much easier because we no longer need to distinguish $S^{\enu}$ and $S^{\nenu}$; one can just imagine that $S^{\enu}$ is the whole configuration.
Notice that the valuation function $\val$ is additive inside each configuration.

We slightly abuse the notation and let $\bx\in[0,1]^M$ be the fractional allocation vector.
We again use $\bx^{\nlg}$ be the vector $\bx$ that is restricted to items in $M^{\nlg}$.
Let $\mu^{\nlg}:=F(\bx^{\nlg})$ be the multilinear extension value for non-large items.
Let $t^{\nlg}:=\val(T^{\nlg})$.
By \ref{property:rounding-lp} and a similar proof for \cref{thm:pipage-rounding}, we have, for any $\lambda<0$
\begin{equation}
E[\val(S^{\nlg})] \leq \frac{e}{e-1} \cdot F(\bx^{\nlg}):=\mu^{\nlg}\leq\E[t^{\nlg}]
\qquad \text{and} \qquad
\E\left[e^{\lambda t^{\nlg}}\right] \leq e^{(e^{\lambda}-1)\mu^{\nlg}} 
\label{equ:upper-integap:concentration}    
\end{equation}
Applying the same worst-case transformations for the input and output distributions, we shall have the following properties again: (\rom{1}) $T^{\lg}$ and $S^{\lg}$ are the same distribution; (\rom{2}) $\val(T^{\lg})=t^{\nlg}\geq 1$; (\rom{3}) there exists a $k\geq 0$ such that the following inequalities hold:
\begin{equation}
\E\left[\val(S^{\nlg})\right]=\E[(k-\val(S^{\lg})^+]
\text{ and }
\E\left[\ln\left(\val(S^{\lg})+\val(S^{\nlg})\right)\right]\leq \E\left[\ln\left(\val(S^{\lg})+(k-\val(S^{\lg}))^+\right)\right]
\label{equ:upper-integap:k}
\end{equation}

Hence, we have a mathematics programs \mpref{mp:five} which is similar to \mpref{mp:three} and \mpref{mp:four}. 
Note that \mpref{mp:five} is essentially equivalent to the special case of \mpref{mp:three} where $\alpha=0$. But we do not need to define $\bar\mu$ in \mpref{mp:five}; so \ref{mp5:k} is not an equation:

\mppara{5}{mp:five}{
Our goal is to maximize
\begin{equation}
\E\left[\left(\ln k-\ln t^{\nlg}\right)^+\right],
\label{equ:upper-integap:ratio}
\end{equation}
subject to the following constraints:
\begin{enumerate}[label=(MP5\alph*),leftmargin=*]
    \item $t^{\nlg}\geq 1$ is a random variable and $k\geq 0$ is a real number.
    \label{mp5:1}
    \item $\E[t^{\nlg}]\geq \mu^{\nlg}$.
    \label{mp5:mu}
    \item For any $\lambda<0$, $\E\left[e^{\lambda t^{\nlg}}\right] \leq e^{(e^{\lambda}-1)\mu^{\nlg}} $.
    \label{mp5:concentration}
    \item $\E[(k-t^{\nlg})^+] \leq \left(\frac{e}{e-1}\right)\cdot \mu^{\nlg}$,
    \label{mp5:k}
\end{enumerate}
}
where \eqref{equ:upper-integap:ratio} is obtained in the same way as \eqref{equ:ratio-4}, \ref{mp5:mu} and \ref{mp5:concentration} are due to \eqref{equ:upper-integap:concentration}, \ref{mp5:k} is due the first inequality of \eqref{equ:upper-integap:concentration} and the first equality of \eqref{equ:upper-integap:k}.

Similar to \cref{subsec:computer-program}, we first show that \eqref{equ:upper-integap:ratio} in \mpref{mp:five} is small when $\mu^{\nlg}$ is very large and very small.
For simplicity, we again use $t$ and $\mu$ to represent $t^{\nlg}$ and $\mu^{\nlg}$.
Same as the concentration bound in \eqref{equ:mp4-concentration-k}, \ref{mp5:mu} and \ref{mp5:concentration} imply the Chernoff-type low-tail bound:
\begin{equation}
\Pr\left[t\leq (1-\delta)\mu\right]\leq e^{-\delta^2\mu/2},\quad \text{for all }\delta\in(0,1).
\label{equ:upper-integap:chernoff}    
\end{equation}

\begin{corollary}
In \mpref{mp:five}, when $\mu\leq 1$, the value of \eqref{equ:upper-integap:ratio} is at most $\ln(1+e/(e-1))$.
\end{corollary}

\begin{lemma}
In \mpref{mp:five}, when $\mu\geq 1000$, the value of \eqref{equ:upper-integap:ratio} is at most $\ln(1+e/(e-1))+0.280409$.
\end{lemma}

\begin{proof}
Apply the same proof as \cref{lem:large-mu} and \cref{lem:large-p:ratio}, \eqref{equ:upper-integap:ratio} is equal to the summation of the following two terms:
\begin{fleqn}[0pt]
    \begin{align}
    &\Pr\left[t\geq (1-\delta)\mu\right]\cdot\E\left[ \ln\left(\frac{(k-t)^+}{t}+1\right)\Bigg|t\geq (1-\delta)\mu\right]\leq \frac{\ln\left(1+\left(\frac{e}{e-1}\right)\mu\right)}{e^{\delta^2\mu/2}}:=y(\mu); \label{equ:upper-integap:right} \\
    &\Pr\left[t\leq (1-\delta)\mu\right]\cdot\E\left[ \ln\left(\frac{(k-t)^+}{t}+1\right)\Bigg|t\leq (1-\delta)\mu\right]\leq \ln\left(1+\frac{e}{e-1}\right)+2\delta, \label{equ:upper-integap:left}
    \end{align}
\end{fleqn}
where \eqref{equ:upper-integap:right} applied \eqref{equ:upper-integap:chernoff}, \ref{mp5:1}, \ref{mp5:k}, and \eqref{equ:upper-integap:left} applied \eqref{equ:upper-integap:chernoff}, \ref{mp5:k}.
We again have the following upper bound on \eqref{equ:upper-integap:ratio}:
\[
\eqref{equ:upper-integap:ratio}=\E\left[\left(\ln k-\ln t\right)^+\right] \leq \ln\left(1+\frac{e}{e-1}\right)+2\delta+y(\mu).
\]
One can verify that the following inequality holds if $\mu\geq 1000$ and $\delta=0.14$ (note again that $y(\mu)$ decreases in $[1000,+\infty)$):
\[
y(\mu)<0.000409
\quad\text{and}\quad
\exp\left(\ln\left(1+\frac{e}{e-1}\right)+y(\mu)+2\delta\right)<3.418.
\]
This finishes proving the lemma.
\end{proof}

Finally, we use the computer program to show that the integrality gap is at most $\ln(3.45)>1.238$. Similar to the previous case, we can use \cref{lem:bound-k} to show that the value of \mpref{mp:five} is small when $\bar\mu\le 1000$ and $k>30000$, So it remains to run the program for parameters $1\le \bar\mu\le 1000,1\le k\le 30000$, which is similar to the one in \cref{subsec:code}.
In the sense of coding, \mpref{mp:five} is indeed a special case of \mpref{mp:three} when $\alpha=0$.
Thus, we only need to change the range of parameters $k$ and $\mu$; see the code in \cref{subsec:code} for details.

\section{Integrality Gap for Submodular Valuations}
\label{sec:submodular-gap}

This section shows that the integrality gap of~\eqref{Conf-LP} is at least $2^{\ln 2}-\delta\approx 1.6168-\delta$ for any $\delta>0$, for weighted NSW with submodular valuations. 
Note that $2^{\ln 2}$ is strictly larger than the current best-known hardness result ${e}/({e-1})\approx 1.5819761$ for NSW with submodular valuations given by~\cite{talg/GargKK23}.
Formally, we aim to show \cref{thm:submodular_gap}.

Our set system is built on a partition system proposed by~\cite{jacm/Feige98,algorithmica/KhotLMM08}, which is used to show the lower bound of the submodular social welfare problem, whose goal is to partition a set of items among agents such that the sum of the agents' utilities is maximized.

\paragraph{The gap instance $\cI$.}
Let $2 \leq \lambda < k$ be two integers whose values will be decided later. Let $h = k\lambda$ and $r = k^h$. Let $\epsilon > 0$ be sufficiently small, and $t > 0$ be a sufficiently large value.  
The set of items is defined as follows:
\begin{itemize}
    \item There are $hk = k^2\lambda$ \emph{set items}, each correspondent to a subset of the ground set $[k]^h$ of size $r$; so $r=k^h$. For every $p \in [k], q \in [\lambda], o \in [k]$, we define the item $A^{p, q}_o$ to be $\{v \in [k]^h: v_{(p-1)\lambda + q} = o\}$.
    Thus $(A^{p, q}_o)_{o \in [k]}$ for any $p \in [k], q \in [\lambda]$ is a partition of the grid $[k]^h$ using the $((p-1)\lambda + q)$-th coordinate.
    \item There are $k - \lambda$ \emph{large items}.
\end{itemize}

We then define the set of agents. There are $k$ groups of agents $N^1, N^2, \ldots, N^k$. Each group $N^p, p \in [k]$ contains a heavy agent $i^{\hv}_p$ and $\lambda (k-1)$ light agents $\Big\{i^{\lt}_{p, q, o}: q \in [\lambda], o \in [k-1]\Big\}$. The heavy agent $i^\hv_p$ has weight $\frac{1-\epsilon}{k}$, and each light agent $i^\lt_{p, q, o}$ has weight $\frac{\epsilon}{k\lambda(k-1)}$. So, the total weight of agents in each group $N^p$ is $\frac{1-\epsilon}{k} + \frac{\epsilon}{k\lambda(k-1)}\cdot \lambda(k-1) = \frac{1-\epsilon}{k} +  \frac{\epsilon}{k} = \frac1k$. The total weight of all agents is 1. Then, we define the valuation functions. 
\begin{itemize}
    \item Focus on a heavy agent $i^\hv_p, p \in [k]$. First, we consider the family of set items of the form $A^{p, q}_o, q \in [\lambda], o \in [k]$ that are assigned to the agent. He gets a value equaling the size of the union of these sets. Then, if he gets at least one large item, he gets an additional value of $t$.
    
    \item Focus on each light agent $i^\lt_{p, q, o}$, $p \in [k], q \in [\lambda], o \in [k-1]$. His value is 1 if he gets at least one set item of the form $A^{p, q}_{o'}, o' \in [k]$, and 0 otherwise. 
\end{itemize}

\begin{figure}[htb]
    \centering
    \begin{subfigure}[t]{0.45\textwidth}
        \includegraphics[width=\linewidth,page=1]{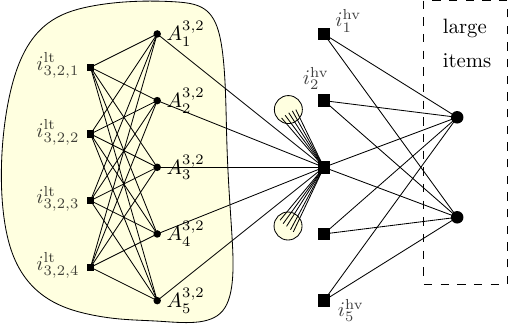}
        \caption{A line between an item and an agent means the item has a positive value to the agent. The big yellow body contains the set items $\{A^{3, 2}_o:o \in [k]\}$ and light agents $\{i^\lt_{3, 2, o}:o \in [k-1]\}$. There are $\gamma \times k = 15$ such bodies.}
    \end{subfigure}\hfill
    \begin{subfigure}[t]{0.45\textwidth}
        \includegraphics[width=\linewidth,page=2]{Figures/submodular_gap_frac_new-1.pdf}      
        \caption{Each $i^{\lt}_{p, q, o}$ is assigned to each $i^{\lt}_{p, q, o'}$ with fraction $\frac1k = \frac15$ as a singleton configuration. $i^\hv_p$ gets $\frac15$ fraction of the configuration $\{A^{p, q}_o: o \in [k]\}$ for every $p, q$. Every $i^\hv_p$ gets a $\frac15$ fraction of every large item as a singleton configuration.}
    \end{subfigure}

    \caption{Illustration of the gap instance to \eqref{Conf-LP} with $k = 5$ and $\lambda = 3$. Big and small squares denote the heavy and light agents, respectively, and big and small circles denote the large and small items, respectively.
    }
    \label{fig:submodular-gap:frac}
\end{figure}

Clearly, all valuation functions are coverage functions and thus submodular.
The instance is shown in \cref{fig:submodular-gap:frac}(a).

In the following, we bound the optimal integral and fractional value in \cref{lem:submodular_gap:integral} and \cref{lem:submodular_gap:fractional}, respectively.
Combining these two bounds proves \cref{thm:submodular_gap}.
\begin{lemma}
The optimum value of $\cI$ is at most
\[
\iopt \leq \left(t+r\right)^{(1-\epsilon)\frac{k-\lambda}{k}}\cdot \left( r\left( 1-\left( 1-\frac{1}{k} \right)^{\lambda} \right)\right)^{(1-\epsilon)\frac{\lambda}{k}}.
\]
\label{lem:submodular_gap:integral}
\end{lemma}

\begin{proof}
    
It is not hard to see that the following allocation is optimal. Each light agent $i^\lt_{p, q, o}$ gets the item $A^{p, q}_o$ and thus value 1. This will leave the item $A^{p, q}_k$ unassigned, for every $p \in [k], q \in [\lambda]$. We assign the items $\{A^{p, q}_k, q \in [\lambda]\}$ to $i^\hv_p$. The value from the set items assigned to the heavy agent $i^\hv_p$ is the size of the union of the sets, which is
\begin{align*}
    r \left( 1-\left( 1-\frac{1}{k} \right)^{\lambda} \right).
\end{align*} 
There are $k - \lambda$ heavy agents who will get a large item. This finishes the proof of \cref{lem:submodular_gap:integral}.
\end{proof}

\begin{lemma}
\label{lem:submodular_gap:fractional}
For the instance $\cI$, the exponential $\fopt$ of the optimum value of \eqref{Conf-LP} is at least
\[
\fopt \geq r^{(1-\epsilon)\frac{\lambda}{k}}\cdot t^{(1-\epsilon)\frac{k-\lambda}{k}}.
\]
\end{lemma}

\begin{proof}
Consider the following fractional solution to \eqref{Conf-LP}; See \cref{fig:submodular-gap:frac}(b) for an illustration. 
\begin{itemize}
    \item Each heavy agent $i^\hv_p, p \in [k]$ gets $\frac{k - \lambda}{k}$ fraction of large configurations, where each configuration contains a single large item. As there are $k$ heavy agents and $k-\lambda$ large items, the assignment can be made. Obviously, each configuration has a value of $t$.
    \item Focus on some $p \in [k]$, and we shall describe how to assign the remaining $\frac\lambda k$ fraction of configurations for $i^\hv_p$. The heavy agent $i^\hv_p$ will get $1/k$ fraction of the configuration $\{A^{p, q}_o:o \in [k]\}$ for every $q \in [\lambda]$. So, each configuration has value $r = k^h$, as it is a partition of the ground set $[k]^h$. 
    \item So, $\frac{k-1}{k}$ fraction of each set item $A^{p, q}_o, p \in [k], q \in [\lambda], o \in [k]$ is unassigned.  Focus on each $p$ and $q$. The fractional parts in $\{A^{p, q}_o:o \in [k]\}$ that are unassigned is $k \cdot \frac{k-1}{k} = k- 1$. We can clearly assign them to the $k-1$ light agents $\{i^\lt_{p, q, o}:o \in [k-1]\}$. Each configuration assigned to a light agent has a value of 1.
\end{itemize}

In summary, we have
\begin{align*}
\ln(\fopt) 
&\geq k\cdot \frac{1-\epsilon}{k}\cdot\left(\frac{\lambda}{k}\ln r+\frac{k-\lambda}{k}\ln t\right) = (1-\epsilon)\left(\frac{\lambda}{k}\ln r+\frac{k-\lambda}{k}\ln t\right).
\end{align*}
This finishes the proof of \cref{lem:submodular_gap:fractional}.
\end{proof}

\begin{proof}[Proof of \cref{thm:submodular_gap}]
Combing \cref{lem:submodular_gap:fractional} and \cref{lem:submodular_gap:integral}, we have
\begin{align*}
\lim_{t \to \infty} \lim_{\epsilon \to \infty}\frac{\fopt}{\iopt} 
&\geq \lim_{t \to \infty} \lim_{\epsilon \to \infty}\left( \frac{t}{t+r} \right)^{\frac{(1-\epsilon)(k-\lambda)}{k}}  \cdot \left( 1- \left(1-\frac{1}{k} \right)^{\lambda}\right)^{-\frac{(1-\epsilon)\lambda}{k}} \\
&= \lim_{t \to \infty} \left( \frac{t}{t+r} \right)^{\frac{k-\lambda}{k}} \cdot \left( 1- \left(1-\frac{1}{k} \right)^{\lambda}\right)^{-\frac{\lambda}{k}} \\
&=\left( 1- \left(1-\frac{1}{k} \right)^{\lambda}\right)^{-\frac{\lambda}{k}}
\end{align*}
We let $k$ tend to $\infty$, and keep $\lambda = \floor{ck}$ for a constant $c \in (0, 1)$. The above bound will tend to $\left( 1-\frac{1}{e^c} \right)^{-c}$. 
The quantity gets its maximum value $2^{\ln 2}$ at $c := \ln 2$. Therefore, if we let $k$ be sufficiently large, $\gamma = \floor{ck}, h = \gamma k, r = k^h$, $t$ be sufficiently large depending on $r$, and $\epsilon$ to be small enough depending on all previous parameters, then the gap can be made arbitrarily close to $2^{\ln 2}$.  
This finishes the proof of \cref{thm:submodular_gap}.
\end{proof}

\section{Integrality Gap for Additive Valuations}

This section shows that the integrality gap of \eqref{Conf-LP} is $e^{1/e}-\delta$ for any constant $\delta>0$ when valuation functions are additive and agents are weighted. 
So, the $e^{1/e} + \epsilon$ approximation ratio given by \cite{icalp/FengLi24} is tight. In a restricted assignment instance, every item $j \in M$ has a value $v_j$, and for every $i \in N$, we have $v_i(j) \in \{0, v_j\}$.

\paragraph{The gap instance $\cI$.}
Let $k<h$ be two integers, which later we will let $\frac{k}{h}$ approach $1-\frac{1}{e}$. Let $\epsilon>0$ be a sufficiently small constant, and let $t>0$ be a sufficiently large value.
We first define the agent set.
The agent set $N$ contains $h$ groups of agents: $N^1,\ldots,N^h$.
Fix a group index $p$, we have:
\begin{itemize}
    \item The agent group $N^p$ includes 1 heavy agent $i^{\hv}_p$, which has a weight of $\frac{1-\epsilon}{h}$.
    \item The agent group $N^p$ includes $k$ light agents $i^{\lt}_{p,1},\ldots,i^{\lt}_{p,k}$, each of which has a weight of $\frac{\epsilon}{kh}$. 
\end{itemize}
Hence, the total weight of heavy agents is $h\cdot\frac{1-\epsilon}{h}=1-\epsilon$, and the total weight of light agents is $kh\cdot\frac{\epsilon}{kh}=\epsilon$.
So, the total weight of all agents is 1.

The item set includes two types: small and large items, denoted by $M^\sm,M^{\lg}$.
The small item set includes $h$ groups $M^\sm_1,\ldots,M^\sm_{h}$, each with $h$ items; so, $\abs{M^\sm}=h^2$.
The large item set contains $k$ items. 
Fix an agent group $N^p$; we define the valuation functions as follows.
\begin{itemize}
    \item For the heavy agent $i^{\hv}_p$, each item in $M^{\lg}$ has a value of $t$ to this agent. Only small items in $M^{\sm}_p$ have a value of 1 to this agent.
    \item For each light agent $i^{\lt}_{p,q},p \in [h], q\in[k]$, small items in $M^\sm_p$ have a value of 1; other items have value $0$. 
\end{itemize}

\begin{figure}[htb]
    \begin{subfigure}[t]{0.45\textwidth}
        \includegraphics[width=\linewidth,page=1]{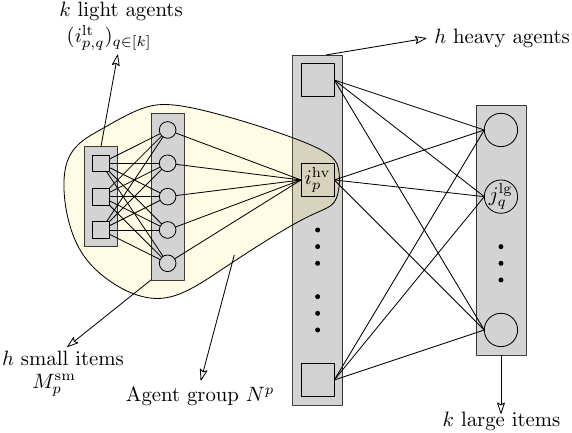}
        \caption{The gap instance.}
    \end{subfigure}\hfill
    \begin{subfigure}[t]{0.45\textwidth}
        \includegraphics[width=\linewidth,page=2]{Figures/additive_gap.pdf}
        \caption{The fractional solution.}
    \end{subfigure}
    \centering
    \caption{Illustration for the gap instance to \eqref{Conf-LP} when the valuation function is additive. The large and small rectangles represent the heavy and light agents, respectively. The large and small circles represent the large and small items, which have values $t$ and $1$, respectively. For each heavy agent $i^{\hv}_p$, there is a group of private light agents $(i^{\lt}_{p,q})_{q\in[k]}$ and small items $M^\sm_p$. A line between an agent and an item indicates the item can be assigned to the agent (with a non-zero value).
}
    \label{fig:additive-gap}
\end{figure}

Clearly, the instance $\cI$ is a restricted assignment instance.
The instance is shown in \cref{fig:additive-gap}.

We bound the optimal integral and fractional value in \cref{lem:additive-gap:iopt} and \cref{lem:additive-gap:fopt}, respectively.
Combining these two bounds proves \cref{thm:additive-gap}.

\begin{lemma}
The optimum value of $\cI$ is
\[
\iopt = \left( (t + h - k)^{\frac{k}{h}} \cdot (h-k)^{\frac{h-k}{h}} \right)^{1-\epsilon}.
\]
\label{lem:additive-gap:iopt}
\end{lemma}
\begin{proof}
It is not hard to see that the following assignment is the optimal integral solution.
$k$ large items are assigned to $k$ different heavy agents. 
Additionally, each of the $h$ heavy agents gets $h-k$ small items. 
Each light agent receives a single small item, and it is assigned a value of 1.
Thus, we have
\begin{align*}
\iopt = 1^{\frac{\epsilon}{kh}\cdot kh} \cdot (t + h - k)^{\frac{1-\epsilon}{h}\cdot k} \cdot \left( h-k \right)^{\frac{1-\epsilon}{h}(h-k)}.   
\end{align*}
This proves \cref{lem:additive-gap:iopt}.
\end{proof}

\begin{lemma}
For the instance $\cI$, the exponential $\fopt$ of the optimum value of \eqref{Conf-LP} is at least:
\[
\fopt \geq (t^{\frac{k}{h}} \cdot h^{\frac{h-k}{h}})^{1-\epsilon}.
\]
\label{lem:additive-gap:fopt}
\end{lemma}

\begin{proof}
Consider the following fractional solution to \eqref{Conf-LP}, which is similar to the proof of \cref{lem:submodular_gap:fractional}. See Figure~\ref{fig:additive-gap}(b) for illustration. 
The assignment is symmetric among all agent groups, so we focus on one agent group, consisting of $i^{\hv}_p,i^{\lt}_{p,1},\ldots,i^{\lt}_{p,k}$.
We describe how the items in $M^{\lg}$ and $M^\sm_p$ are distributed.
\begin{itemize}
    \item The heavy agent $i^{\hv}_{p},p\in[h]$ gets $\frac{k}{h}$ fractions of the configuration of large items, where each configuration contains a single large item. Each configuration has a value of $t$.
    \item The heavy agent $i^{\hv}_{p}$ will also get $1-\frac{k}{h}$ fraction of the configuration $M^\sm_p$, whose value is $h$ since $\abs{M^\sm_p}=h$. So, each light item has a fraction of $\frac{k}{h}$ remaining.
    \item Each light agent $i^{\lt}_{p,q}$ gets $\frac{1}{h}$ fraction of each configuration that includes a single small item; so each configuration has a value of 1. The light agent gets one configuration as there are $h$ small items in $M^\sm_p$. There are $k$ light agents, and they take $\frac{k}{h}$ fractions of each small item in total, so the assignment can be made. 
\end{itemize}
In summary, we have
\begin{align*}
\ln (\fopt) \geq \frac{\epsilon}{kh}\cdot kh \cdot \ln (1) + \frac{1-\epsilon}{h} \cdot h \left( \frac{k}{h}\ln t + (1-\frac{k}{h})\ln k \right).
\end{align*}
This finishes the proof of \cref{lem:additive-gap:fopt}.
\end{proof}

\begin{proof}[Proof of \cref{thm:additive-gap}]
Combining \cref{lem:additive-gap:iopt} and \cref{lem:additive-gap:fopt}, we have
\begin{align*}
\frac{\fopt}{\iopt} \geq \left(\frac{t}{t+h-k}\right)^{\frac kh(1-\epsilon)}\cdot\left(\frac{h}{h-k}\right)^{\big(1-\frac{k}{h}\big)(1-\epsilon)}.
\end{align*}
We let $h$ tend to $\infty$, $k = \floor{(1-1/e)h}$, $t$ tend to $\infty$ depending on $h$ and $k$, and $\epsilon$ tend to $0$. The quantity can be made arbitrarily close to $e^{1/e}$.
\end{proof}

\section{Conclusion}
\label{sec:conclusion}

In this paper, we studied the Nash social welfare problem with submodular valuations.
For weighted NSW, we obtain a $(3.56+\epsilon)$-approximation, simultaneously improving the previous best-known $(233+\epsilon)$-approximation for weighted NSW and $(4+\epsilon)$-approximation for unweighted NSW. 
On the negative side, we show that the configuration LP has an integrality gap $(2^{\ln 2}-\epsilon)$ for weighted NSW with submodular valuations, and $(e^{1/e}-\epsilon)$ with additive valuations.
This rules out the possibility of having a better approximation ratio for weighted NSW with additive valuations based on the configuration LP.

Our work leaves several interesting future directions.
Firstly, we think that our rounding algorithm can achieve a clear $1+(e/(e-1))^2+\epsilon\approx 3.503+\epsilon$ approximation.
Our current analysis can only achieve this ratio when the non-large items are well-concentrated, while we still need to lose a small factor ($\approx 0.06$) on concentration when non-large items are not well-concentrated.
We remark that the value of \mpref{mp:three} is strictly larger than $\ln(1+(e/(e-1))^2)$, and thus, our current analysis cannot achieve this ratio in general.
It would be interesting to see an analysis of $1+(e/(e-1))^2+\epsilon$, especially without the help of the computer program.

On the negative side, we think that the approximation of $1+(e/(e-1))^2+\epsilon$ is a barrier to our algorithmic framework (solving configuration LP plus a matching-pipage style rounding).
Because our framework has to lose a factor $1$ of large items.
For small items, we have to lose one $e/(e-1)$ factor by solving the configuration LP and another $e/(e-1)$ factor by applying the concentration bound; one can just imagine the second $e/(e-1)$-factor comes from the gap between the concave and multilinear extension of a submodular function.
Thus, it would be very interesting to see any algorithm that beats the ratio of $1+(e/(e-1))^2$ even in the unweighted case.

Last but not least, the major part of our integrality gap instance heavily depends on the weights of agents, so they do not hold for unweighted cases.
Currently, the best approximation for weighted NSW with both additive and submodular valuations is achieved by configuration-LP-based rounding.
This paper presents an improved approximation algorithm for the unweighted case, so the best known approximation ratio for unweighted NSW with submodular valuations is also achieved by a configuration-LP-based rounding.
But both analyses are per-agent analyses, and this style of analysis works well for weighted agents, but cannot utilize the advantage of the unweighted agent case.  
Hence, it would be interesting to see whether the configuration LP can be used to beat $(e^{1/e}+\epsilon)$-approximation for unweighted NSW with additive valuations.

\clearpage
\newpage
\bibliographystyle{plain}
\bibliography{main}

\clearpage
\newpage
\appendix
\section{Missing Proofs from \cref{sec:preliminaries}}

\subsection{Proofs for Greedy Proxy Function}
\label{app:greedy-proxy}

This section aims to prove \cref{lem:greedy_proxy}.
We note that $\val(\cdot)$ can be computed in polynomial time.
This is because, given any set $S$, computing the value of $\val(S)$ is equivalent to minimizing the function $h(T):=v(T)-\sum_{j\in T}\phi_j$ over all subsets in $S$.
It is easy to see that $h$ is submodular but not necessarily monotone.
Thus, for any $S\subseteq U$, the value of $\val(S)$ can be computed in strongly polynomial time~\cite{jacm/IwataFF01,jct/Schrijver00}.
We also note that \ref{prop:marginal_value} together with \ref{prop:additive} implies that $\val(P_i)=v(P_i)$ for all $i\in[k]$.

\begin{proofof}{\cref{lem:greedy_proxy}}
We prove that the function $v$ defined in \cref{def:greedy_proxy} satisfies all the properties of \cref{lem:greedy_proxy}.
We prove those properties one by one.
Recall that $\pi^i(t)\in P_i$ is the element at the $t$-th position and $\pi^i([t])\subseteq P_i$ is the first $t$ elements ($\pi^i([0])=\emptyset$).

\paragraph{Property \ref{prop:monotone_submodular}.}
It is easy to see that $v$ is a non-negative function.
The function $v$ is a convolution of a submodular and modular function, and thus, it is submodular; a proof can be found in~\cite[Proposition B.4]{book/Bach13}.
It remains to show the monotonicity of the function $\val$, i.e., for any set $S\subseteq L\subseteq U$, we need to prove $\val(S)\leq \val(L)$.
Let $O_L\subseteq L$ be a minimizer for $\val(L)$; so, $\val(L)=v(O_L)+\sum_{j\in L\setminus O_L}\phi_j$.
Let $O_S:=O_L\cap S$, and we have $\val(S)\leq v(O_S) + \sum_{j\in S\setminus O_S}\phi_j$ since $O_S\subseteq S$.
To prove the monotonicity, it remains to show
\begin{equation}
v(O_S)+\sum_{j\in S\setminus O_S}\phi_j \leq v(O_L) + \sum_{j\in L\setminus O_L} \phi_j.
\label{equ:constructed_sub:monotone}
\end{equation}
Since $f$ is a monotone function, we have $v(O_S)\leq v(O_L)$ because of $O_S\subseteq O_L$.
And, we have $\sum_{j\in S\setminus O_S}\phi_j \leq \sum_{j\in L\setminus O_L} \phi_j$ because $(S\setminus O_S)\subseteq (L\setminus O_L)$ and $\phi_j\geq 0$ for all $j\in U$.
Hence, \eqref{equ:constructed_sub:monotone} holds, which implies the monotonicity of the function $\val$.

\paragraph{Property \ref{prop:lower_bound}.}
For each element set $S\subseteq U$, by choosing $T=S$, we have $\val(S)\leq v(S)$.

\paragraph{Property \ref{prop:marginal_value}.}
Consider an arbitrary element $j\in P_i$ and assuming $j$ is at the $t$-th position.
The value of $\val(j)$ is the minimization over $T\subseteq \set{j}$; so $T$ can be either $\set{j}$ or $\emptyset$.
\begin{itemize}
    \item If $T=\emptyset$, then $v(T)+\sum_{j\in \set{j}\setminus T}\phi_j=\phi_j=v(\pi^i([t]))-v(\pi^i([t-1]))$.
    \item If $T=\set{j}$, then $v(T)+\sum_{j\in \set{j}\setminus T}\phi_j=v(j)$.
\end{itemize}
Since $f$ is a submodular function, we have $v(\pi^i([j-1])\cup\set{j})-v(\pi^i([j-1]))\leq v(j)$.
Thus, the set $T=\emptyset$ is the minimizer and we have $\val(j)=\phi_j=v(\pi^i([t]))-v(\pi^i([t-1]))$.

\paragraph{Property \ref{prop:additive}.}
This property implies that the function $\val$ is an additive function inside each partition $P_i$.
Fix an arbitrary $i\in[k]$ and $S\subseteq P_i$, by \ref{prop:monotone_submodular}, we know that $\val$ is a submodular function.
Thus, we have $\val(S)\leq\sum_{j\in S}\val(j)$.
Hence, to prove \ref{prop:additive}, it is sufficient to show $v(T)+\sum_{j\in S\setminus T}\phi_j\geq \sum_{j\in S}\phi_j$ for any $T\subseteq S$.
Since $\sum_{j\in S}\phi_j=\sum_{j\in T}\phi_j+\sum_{j\in S\setminus T}\phi_j$, it turns out that it is sufficient to prove $v(T)\geq\sum_{j\in T}\phi_j$ for any $T\subseteq S$.
Suppose that $T=\set{j_{\ell_1},\ldots,j_{\ell_r}}$ where elements follows the order in $\pi^i$, i.e., the element $j_{\ell_k}$ is at the $\ell_k$-th position in the order $\pi^i$.
For the index $k=0,1\ldots,r$, we define $A_{k}:=\pi^i([\ell_{k}])\cap T$, where $A_0=\emptyset$ assuming $\pi^i([\ell_0])=\emptyset$.
Since $A_k\subseteq \pi^{i}([\ell_k])$, by submodularity, we have the following inequality for each $k=1,\ldots,r$,
\[
\phi_{j_{\ell_k}}=v(\pi^i([\ell_{k-1}])\cup\set{j_{\ell_k}})-v(\pi^i([\ell_{k-1}])) \leq v(A_{k-1}\cup\set{j_{\ell_k}}) -v(A_{k-1}).
\]
Summing over $k=1$ to $r$, we have
\[
\sum_{k=1}^{r}\phi_{j_{\ell_k}}=\sum_{j\in T}\phi_j\leq v(T)-v(\emptyset)=v(T).
\]
This proves the additivity of $\val$ inside each $P_i,i\in[k]$.
\end{proofof}

\subsection{Proofs for Concentration Bound}
\label{app:concentration}

This section aims to prove the concentration bound stated in \cref{thm:pipage-rounding}.
As~\cite{stoc/FHLZ25,soda/HarveyO14}, we still need to utilize the technique of {\em concave pessimistic estimator}.
Let $\bx,\bx^\smm\in[0,1]^{\bar n}$, $v\in 2^{\bar n}\to\R_{\geq 0}$, $F:[0,1]^{\bar n}\to\R_{\geq 0}$, $\mu:=F(\bx)$, $\mu^\smm:=F(\bx^\smm)$, $S$, $\tau$, and $\bx^{\inte}$ be as defined in the theorem.
Let $\cP(\bx)$ be the product distribution over $2^{\bar n}$, where $\Pr_{V\sim\cP(\bx)}[i\in V]=x_i$ for each $i\in[\bar n]$.
We use the same concave pessimistic estimator as~\cite{soda/HarveyO14}:
\[
g_{t,\theta}(\bx):=e^{-\theta t} \cdot \E_{V\sim \cP(\bx)}\left[e^{\theta v(V)}\right],\quad \forall \bx\in[0,1]^{\bar n},
\]
where $\delta\in(0,1)$ is a fixed constant, $t:=(1-\delta)\cdot\mu$ and $\theta:=\ln(1-\delta)<0$.
The concave pessimistic estimator has a nice property (\cref{lem:estimator-concave}): the expected value of the concave pessimistic estimator $g_{t,\theta}$ does not increase as \cref{alg:pipage} runs; the proof can be found in~\cite[Lemma A.2]{stoc/FHLZ25}.
\cref{lem:estimator-concave} is essentially due to the fact that $g_{t,\theta}$ is concave on the direction $(\be_a)$ and $(\be_a-\be_b)$ for any $a,b\in[\bar n]$, which was proved by~\cite{soda/HarveyO14}.
Note that \cref{lem:estimator-concave} requires  $\theta<0$.

\begin{lemma}[\cite{stoc/FHLZ25}]
Let $\cA(\bx)$ be the output distribution of \cref{alg:pipage} when the input vector is $\bx\in[0,1]^{\bar n}$.
Then, we have $\E_{\bx^{\inte}\sim \cA(\bx)}[g_{t,\theta}(\bx^{\inte})]\leq g_{t,\theta}(\bx)$.
\label{lem:estimator-concave}
\end{lemma}

Besides the concave pessimistic estimator, we need some other tools to \cref{thm:pipage-rounding}.
Let $X_1,\ldots,X_{\bar n}\in\set{0,1}$ be the independent random variables, where $\Pr[X_i=1]=x_i$.
We slightly abuse the notation and we also use $v(X_1,\ldots,X_{\bar n})$ to denote the value of random set $R$ with $R:=\set{i\in[\bar n]:X_i=1}$.
We define the random variable $Y_i$ as follows, which intuitively transforms the submodular function into ``additive'' so that we can bound the exponential moment for submodular functions:
\begin{equation}
Y_i:=v(X_1,\ldots,X_i,0,\ldots,0)-v(X_1,\ldots,X_{i-1},0,\ldots,0).
\label{equ:Y_i}
\end{equation}
By the definition of $Y_i$, we have $v(X_1,\ldots,X_{\bar n})=\sum_{i=1}^{\bar n} Y_i$ and $Y_i\leq 1$ for each $i\in[\bar n]$.
Note that the random variables $Y_i$ are no longer independent.
Fortunately, they still have some negative correlation properties (\cref{lem:negative-correlation}), which were shown in~\cite{focs/ChekuriVZ10}; the proof can be found in the full version~\cite[Lemma C.1.]{arxiv/abs-0909-4348} of \cite{focs/ChekuriVZ10}.

\begin{lemma}[\cite{focs/ChekuriVZ10}]
For any $\lambda\in\R$, we have:
$\E\left[e^{\lambda\sum_{i=1}^{\bar n}Y_i}\right] \leq \prod_{i=1}^{\bar n}\E\left[e^{\lambda Y_i}\right].$
\label{lem:negative-correlation}
\end{lemma}

Now, we are ready to prove \cref{thm:pipage-rounding}.

\begin{proofof}{\cref{thm:pipage-rounding}}
Recall that $S$ is the small item set defined in the theorem and $S=\set{i\in[\bar n]:v(i)\leq \tau}$.
Let $L:=[\bar n]\setminus S$ be the remaining large items.
Without loss of generality, we assume that small items are the first $\ell$ items in $[\bar n]$.
Let $\bx^{\lgg}$ be the vector $\bx$ that is restricted to large items.
Let $\bx^{\smm,\inte}$ and $\bx^{\lgg,\inte}$ be the integral ($\set{0,1}$) vectors returned by \cref{alg:pipage} when the input is $\bx^{\smm}$ and $\bx^{\lgg}$, respectively. 
Note that $\bx^{\smm,\inte}$ and $\bx^{\lgg,\inte}$ are equivalent to two sets, i.e., $U^{\smm}:=\set{i\in[\bar n]:x^{\smm,\inte}_i=1}$ and $U^{\lgg}:=\set{i\in[\bar n]: x^{\lgg,\inte}_i=1}$.

Recall that $X_1,\ldots,X_{\bar n}$ are the independent random variables with $\Pr[X_i=1]=x_i$ and $Y_i$ is defined as the marginal gain (see \eqref{equ:Y_i}).
Note that $Y_i\leq \tau$ for all $i\in[\ell]$ and $Y_i\leq 1$ for all $i\in[\bar n]$.
Thus, for any $\theta\in\R$, we have
\begin{equation}
\E_{V^{\smm}\sim\cP(\bx^{\smm})}\left[e^{\theta v(V^{\smm})}\right] 
\stackrel{\text{(\rom{1})}}{\leq}
\prod_{i=1}^{\ell}\E\left[ e^{\theta Y_i}\right]
=
\prod_{i=1}^{\ell}\E\left[ e^{\theta\tau \frac{Y_i}{\tau}}\right]
\stackrel{\text{(\rom{2})}}{\leq}
\prod_{i=1}^{\ell}e^{\frac{\E[Y_i]}{\tau}(e^{\theta\tau}-1)}
=
e^{(e^{\theta\tau}-1)\mu^{\smm}/\tau},  
\label{equ:concentration-small}
\end{equation}
where the inequality (\rom{1}) is due to \cref{lem:negative-correlation} and the inequality (\rom{2}) is due to the standard inequalities from the proof of Chernoff bound, as well as $Y_i\leq 1$:
$e^{ab} \leq 1 + (e^a-1) \cdot x \leq e^{(e^a-1)b}, \forall a\in\R, b\in[0,1]$.
Similar to \eqref{equ:concentration-small}, we have the exponential moment for large items:
\begin{equation}
\E_{V^{\lgg}\sim\cP(\bx^{\lgg})}\left[e^{\theta v(V^{\lgg})}\right] 
{\leq}
\prod_{i=\ell+1}^{\bar n}\E\left[ e^{\theta Y_i}\right]
=
\prod_{i=\ell+1}^{\bar n}\E\left[ e^{\theta Y_i}\right]
{\leq}
\prod_{i=\ell+1}^{\bar n}e^{\E[Y_i](e^{\theta}-1)}
\stackrel{\text{(\rom{3})}}{=}
e^{(e^{\theta}-1)(\mu-\mu^{\smm})},  
\label{equ:concentration-large}
\end{equation}
where the equality (\rom{3}) is due to the definition of $Y_i$: 
\[
\sum_{i=\ell+1}^{\bar n}\E[Y_i]=\E[v(X_1,\ldots,X_{\bar n})]-\E[v(X_1,\ldots,X_{\ell},0,\ldots, 0)]=\mu-\mu^{\smm}.
\]
Thus, by merging \eqref{equ:concentration-small} and \eqref{equ:concentration-large}, we have
\begin{equation}
\prod_{i=1}^{\bar n}\E\left[e^{\theta Y_i}\right]=\left(\prod_{i=1}^{\ell}\E\left[e^{\theta Y_i}\right]\right)\cdot\left(\prod_{i=\ell+1}^{\bar n}\E\left[e^{\theta Y_i}\right]\right) \leq e^{(e^{\theta\tau}-1)\mu^{\smm}/\tau+(e^{\theta}-1)(\mu-\mu^{\smm})}.
\label{equ:concentration-independent}
\end{equation}
Note that \eqref{equ:concentration-independent} holds for any $\theta\in\R$ because $X_1,\ldots,X_{\bar n}$ are independent random variables.

It now remains to establish the concentration bound for variables produced by \cref{alg:pipage}, and this is the place where we apply the concave pessimistic estimator.
By \cref{lem:estimator-concave}, \cref{lem:negative-correlation} and the definition of $g_{t,\theta}$, we have
\begin{equation}
\E_{\bx^{\inte}\sim \cA(\bx)}\left[ g_{t,\theta}(\bx^{\inte}) \right]=\frac{\E_{U\sim \cA(\bx)}[e^{\theta v(U)}]}{e^{\theta t}} \leq g_{t,\theta}(\bx)=\frac{\E_{V\sim \cP(\bx)}[e^{\theta v(V)}]}{e^{\theta t}}\leq\frac{\prod_{i=1}^{\bar n}\E\left[e^{\theta Y_i}\right]}{e^{\theta t}},    
\label{equ:concentration-pipage}
\end{equation}
where \cref{lem:negative-correlation} applies to the last inequality due to the fact that $V$ is independently sampled from the product distribution $\cP(\bx)$.
Note that \eqref{equ:concentration-pipage} requires that $\theta<0$.
Now, by merging \eqref{equ:concentration-independent} and \eqref{equ:concentration-pipage}, we have
\[
\E_{U\sim\cA(\bx)}\left[e^{\theta v(U)}\right] \leq \prod_{i=1}^{\bar n} \E\left[e^{\theta Y_i}\right]\leq e^{(e^{\theta\tau}-1)\mu^{\smm}/\tau+(e^{\theta}-1)(\mu-\mu^{\smm})}, \quad \forall \theta < 0,
\]
where the first inequality is due to \eqref{equ:concentration-pipage} and $e^{\theta t}>0$ and the second inequality is due to \eqref{equ:concentration-independent}.
This finishes proving the theorem.
\end{proofof}

\section{Missing Proofs from \cref{sec:alg}}

\subsection{Proofs for LP Solver}
\label{app:lp-solver}

This section aims to prove \cref{lem:lp-solver}, which uses \cref{lem:submodular-cover} as an approximate separation oracle.
Applying a similar method as~\cite{stoc/FHLZ25,icalp/FengLi24}, we can obtain the desired fractional solution in polynomial time via the ellipsoid method.

\begin{proposition}
Given a monotone submodular function $f:2^{U}\to\R_{\geq 0}$ defined on a set of ground elements $U$ with $\abs{U}=n$, an item cost vector $\bc\in\Q_{>0}^n$, and a target value $V\geq 0$.
Let $\opt:=\min_{S\subseteq U:f(S)\geq V}\sum_{i\in S}c_i$ and $O$ be an item set that achieves $\opt$.
Let $\pi$ be the greedy order of elements in $O$ under $f$.
Then, for any constant $\epsilon>0$, there is a polynomial time algorithm that finds a set $S$ such that 
\[
\sum_{i\in S}c_i\leq \opt \quad \text{ and } \quad 
f(S) \geq (1-\epsilon)f(O^{\enu})+\left(1-\frac{1}{e}\right)\cdot \left( f(O)-f(O^{\enu}) \right),
\]
where $O^{\enu}$ is a set of the first $1/\epsilon$ elements on the greedy order $\pi$ if $\abs{O}>1/\epsilon$; otherwise $O^{\enu}=O$.
\label{lem:submodular-cover}
\end{proposition}

\cref{lem:submodular-cover} is achieved by a modified greedy algorithm for the submodular knapsack problem~\cite{orl/Sviridenko04}.
The standard greedy for the submodular knapsack first enumerates all sets of items up to a size $3$ and then applies the greedy algorithm (add elements in the order of their densities).
In the modified version, we need to enumerate all sets of elements of size $1/\epsilon$ and then grow the solution via greedy.
We can always ensure that $c_i$ is integral by scaling.  
If the scaled $c_i$ is not polynomially bounded, the standard binary search technique can be used to handle this case.
In~[Last inequality on last page]\cite{orl/Sviridenko04}, they implicitly proved the following inequality:
\[
f(S)\geq \left(1-\frac{1}{k}\right)f(T)+\left(1-\frac{1}{e}\right)\cdot \left(f(O^*)-f(T)\right),
\]
where $S$ is the solution returned by the greedy algorithm, $O^*$ is the optimal solution to the submodular knapsack problem, and $\abs{T}=k$ is part of the optimal solution that is also enumerated by the algorithm.
By setting $k=1/\epsilon$, this directly proves \cref{lem:submodular-cover}.

\begin{proofof}{\cref{lem:lp-solver}}
The proof is almost the same as the one in~\cite[Lemma 3.1]{stoc/FHLZ25}. 
We now sketch the main proof ideas.
We can assume we have a number $o$ such that the value of \eqref{Conf-LP} is in $(o,o+\frac{\epsilon}{3}]$ by making $O(\frac{\log m}{\epsilon})$ guesses. 
Then, we set up a dual program \eqref{Dual-LP} of \eqref{Conf-LP}, with the objective replaced by a linear constraint. 
\begin{align}
     &&   & \tag{\text{Dual-LP}} \label{Dual-LP}\\
    &&\sum_{j\in M} \alpha_j + \sum_{i\in N}\beta_i &\leq o, & \label{dual:obj} \\
    &&\sum_{j\in S}\alpha_j+\beta_i &\geq w_i\cdot\ln v_i(S), &\forall i\in N, S\subseteq M \label{dual:const}\\
    &&\alpha_j &\geq 0, &\forall j\in M 
\end{align}
As we have that the value of \eqref{Conf-LP} is in $(o, o + \epsilon]$, \eqref{Dual-LP} is infeasible. 
Given two vectors $\balpha\in\R_{\geq 0}^{M}$ and $\bbeta\in\R^{n}$ with $\sum_{j \in M} \alpha_j + \sum_{i \in N} \beta_i \leq o$, we know \eqref{dual:const} is not satisfied for some $i \in N$ and $S \subseteq M$. 
Our goal is to design an approximate separation oracle for the inequality above.  
We apply \cref{lem:submodular-cover}, with item costs $\alpha_j$ and target value $v(S)$ (the value of $v(S)$ can be guessed via standard techniques, similar to~\cite[Lemma 3.1]{stoc/FHLZ25}). 
Then, we find a set $S'$ such that
\[
\sum_{j\in S'}\alpha_j \leq \sum_{j \in S}\alpha_j \qquad \text{and} \qquad v_i(S') \geq (1-\epsilon)f(S^{\enu})+\left(1-\frac{1}{e}\right)\left(v(S)-v(S^{\enu})\right).
\]
This implies that
\begin{align*}
v_i(S)
&\leq v_i(S^{\enu})+\frac{e}{e-1}\left(v_i(S')-(1-\epsilon)v_i(S^{\enu})\right)\\
&\leq (1+O(\epsilon))\cdot \left(v_i(S^{\enu})+\frac{e}{e-1}\left(v_i(S')-v_i(S^{\enu})\right)\right)    
\end{align*}
Since $(i,S)$ is an infeasible solution to \eqref{Dual-LP}, we have
\begin{align*}
\sum_{j\in S'}\alpha_j+\beta_i 
&\leq \sum_{j\in S}\alpha_j+\beta_i \leq w_i\cdot\ln(v_i(S))\\
&\leq w_i\cdot\ln\left((1+O(\epsilon))\cdot \left(v_i(S^{\enu})+\frac{e}{e-1}\left(v_i(S')-v_i(S^{\enu})\right)\right) \right) \\   
\end{align*}
Based on the same argument of~\cite{stoc/FHLZ25}, we have \cref{lem:lp-solver}.
\end{proofof}

\addtocontents{toc}{\protect\enlargethispage{\baselineskip}}
\section{Integrality Gap for Unweighted Agents with Additive Valuations}
\label{sec:unweighted-additive}

This section shows that \eqref{Conf-LP} has an integrality gap of $2^{1/4}-\epsilon \approx 1.189-\epsilon$ for unweighted additive functions. The gap instance is a restricted assignment instance. Moreover, each item has a non-zero value to exactly two agents. So, we just use an edge-weighted graph over $N$ to denote the instance: an edge $(i, i’)$ between two agents $i$ and $i’$ denotes an item that can only be assigned to $i$ and $i’$. The value of the edge is the value of the item when assigned to $i$ or $i’$.  

\begin{figure}[htb]
    \centering
    \includegraphics[width=0.3\linewidth]{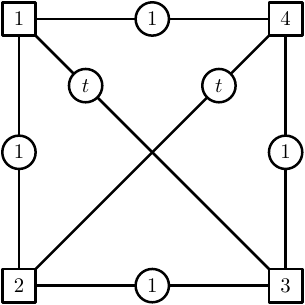}
    \caption{Illustration of gap instance for the unweighted NSW with additive agents. Each rectangle represents an agent, and each circle represents an item. The value inside the rectangle and circle is the agent's index and the item's value, respectively. An agent only has a non-zero value to those items that connect to the agent.}
    \label{fig:unweighted-gap}
\end{figure}

The graph is defined as follows. We have 4 agents indexed as $[4]$. There are 4 edges (1, 2), (2, 3), (3, 4,) and (4, 1) with value 1, and 2 edges (1, 3) and (2, 4) with value $t$, where $t$ tends to $\infty$. So, we can view the 4 small items of value 1 as the 4 sides of a square, and the 2 large items of value $t$ as two diagonals of the square. Recall that all agents have a weight $1/4$. 
See \cref{fig:unweighted-gap} for an example.

Due to the symmetry, we can assume the optimum integral solution assigns the two large (diagonal) items to agents 1 and 2. Then it is best to let agents 1, 2, 3, and 4 get 1, 0, 2, and 1 small (side) items, respectively. The resulting solution has NSW value $((t+1)\cdot t \cdot 2 \cdot 1)^{1/4} = (2t(t+1))^{1/4}$.

Now we describe the solution to \eqref{Conf-LP}. Agent 1 will get $1/2$ fraction of the configuration $\{(1, 3)\}$, and $1/2$ fraction of the configuration $\{(4, 1), (1, 2)\}$. That is, she gets a $1/2$ fractional configuration containing the big item incident to her, and a 1/2 fractional configuration containing the two small items incident to her. The allocation for the other 3 agents can be defined symmetrically.  
This solution has value $\frac12 \ln t + \frac12 \ln2$ to the configuration LP. Thus, we have $\fopt \geq \sqrt{2t}$. 

So, the integrality gap is $\frac{\sqrt{2t}}{(2t(t+1))^{1/4}}$, which approaches $2^{1/4}$ as $t$ tends to $\infty$. 

\end{document}